\documentclass[lineno]{JFM-FLM_Au}

\usepackage{subfigure,color}
\usepackage{comment}
\usepackage{subcaption}
\usepackage{graphicx}
\usepackage{physics}

\newcommand{\bs}[1]{{\boldsymbol{#1}}}

\newcommand{\trans}{{\rm T}}

\newcommand{\sfbiG}{\mathsfbi{G}}
\newcommand{\bsu}{{\bs{u}}}
\newcommand{\bstau}{\bs{\tau}}

\newcommand{\bsq}{\bs{q}}
\newcommand{\bsQ}{\bs{Q}}
\newcommand{\bsK}{\mathsfbi{K}}
\newcommand{\bsU}{\mathsfbi{U}}
\newcommand{\bsV}{\mathsfbi{V}}
\newcommand{\bsW}{\mathsfbi{W}}

\newcommand{\cq}{\check{\bsq}} 
\newcommand{\cQ}{\check{\mathsfbi{Q}}} 
\newcommand{\rSVD}{l} 

\newcommand{\hu}{\hat{\bsu}}
\newcommand{\htau}{\hat{\bstau}}

\newcommand{\tp}{{\breve{p}}}
\newcommand{\tbu}{\breve{\bsu}}

\newcommand{\ttau}{\breve{\bstau}}
\newcommand{\tbG}{\breve{\sfbiG}}

\newcommand{\mP}{\mathcal{P}}

\newcommand{\tnabla}{\tilde{\bnabla}}

\lefttitle{Kun Zhang et al.}
\righttitle{Journal of Fluid Mechanics}

\title{Revisiting creeping viscoelastic cross-slot flow: Global linear stability and structural sensitivity analyses}

\author{Kun Zhang\aff{1}, Zhanwen Wang\aff{1} \and Lailai Zhu\aff{1}}

\affiliation{\aff{1}Department of Mechanical Engineering, National University of Singapore, Singapore 117575, Republic of Singapore}

\corresau{Lailai Zhu, lailai\_zhu@nus.edu.sg}

\begin{document}
\maketitle

\begin{abstract}
The viscoelastic instability of cross-slot flow was first observed experimentally almost half a century ago and reproduced numerically two decades ago, yet its physical origin remains unresolved. We revisit this problem for two-dimensional creeping flow of  Oldroyd-B fluid by combining direct numerical simulations, global stability analysis, structural sensitivity analysis, and energy-budget analysis. Our simulations reproduce the canonical pitchfork bifurcation, and the stability analysis consistently predicts the threshold and perturbation growth rates. The leading eigenmode consists of a chiral velocity--stress perturbation that tilts and rotates the birefringent strand generated by the extensional flow.
Structural sensitivity and energy-budget analyses identify narrow high-extension-rate ridges within the extensional flow as both the spatial core and energetic source of the instability. In these ridges, the stress-based wavemaker co-localizes with large positive disturbance polymeric stress power density, indicating localized transfer of stored elastic energy to the
disturbance flow. Analyses of cross-slot variants with rounded corners and with a centered cylinder further reveal that neither sharp corners nor a free central stagnation point is the essential destabilizing ingredient; rather, the instability originates from elastic-energy release in extension-dominated regions characteristic of cross-slot flow.
\end{abstract}


%{\bf MSC Codes }  {\it(Optional)} Please enter your MSC Codes here

\section{Introduction} \label{sec:Introduction}

The addition of dilute polymers to a Newtonian solvent gives rise to a wide variety of complex flow phenomena~\citep{datta_perspectives_2022}. The resulting non-Newtonian fluid typically possesses viscoelasticity, an essential rheological property underlying the emergence of diverse behaviors such as 
rod-climbing (the Weissenberg effect)~\citep{weissenberg1947continuum,ruangkriengsin2025transient}, die swell~\citep{tanner2005theory,tang2020state}, turbulent drag reduction~\citep{white2008mechanics,graham2014drag,choueiri2018exceeding,wang2023maximum}, and elastic~\citep{groisman2000elastic,steinberg2021elastic,sasmal2025potential} and elasto-inertial turbulence~\citep{samanta2013elasto,zhang2025experimental}. 
A particularly significant subset of these phenomena involves the alteration of hydrodynamic stability by polymer-induced stresses. Unlike Newtonian instabilities, which typically stem from inertial effects~\citep{schmid2001stability}, polymeric flows are influenced by the interplay between inertia and fluid elasticity, leading to the emergence of elasto-inertial instabilities~\citep{dubief2023elasto}. Remarkably, these flows can also become unstable at vanishing Reynolds number via a pure elastic route~\citep{shaqfeh1996purely}. The ensuing elastic instability arises when the elasticity-indicating parameter, the Weissenberg ($Wi$) number (or Deborah number)---ratio of the polymer relaxation time to the characteristic flow convection time---exceeds a threshold.

Extensive studies have analyzed both elastic and elasto-inertial instabilities of viscoelastic flows in canonical settings: Taylor--Couette geometry \citep{muller1989purely,larson1990purely,shaqfeh1992effects,graham1998effect, liu2013elastically, bai2023viscoelastic, beneitez2025linear} and parallel, rectilinear configurations including Kolmogorov \citep{boffetta2005viscoelastic,bistagnino2007nonlinear,berti2010elastic,lewy2025revisiting}, planar Couette \citep{gorodtsov1967linear,renardy1992rigorous,morozov2005subcritical, binns2024global}, planar Poiseuille \citep{grosch1968stability, sureshkumar1995linear, wilson1999structure,khalid2021centre,lellep_purely_2021}, and pipe flows \citep{garg2018viscoelastic, chaudhary_linear_2021,dong2022asymptotic}. 
These geometric configurations admit analytical base-state solutions, which facilitate modal~\citep{renardy1986linear}
and non-modal~\citep{jovanovic2011nonmodal} linear stability analyses, as well as weakly nonlinear analyses \citep{sanchez2022understanding}.
These theoretical studies have provided deep insights into the mechanisms of center-mode instabilities \citep{garg2018viscoelastic}, the energy budgets of the eigenmodes \citep{sadanandan2004global,hoda2008energy,spyridakis_viscoelastic_2024},
and the subcritical nature of the transition in shear-dominated parallel flows \citep{morozov2007introductory,garg2018viscoelastic, buza_weakly_2022}.

Despite extensive stability analyses of viscoelastic flows with analytical base states, far fewer efforts have addressed those without such base states---typically arising in 
non-rectilinear geometries. By contrast, a surprisingly large body of experimental and numerical work has reported unstable viscoelastic flows in a broad range of complex geometric domains, including cross-slot \citep{arratia2006elastic,canossi2020elastic}, pore--throat \citep{de2017viscoelastic,browne2020bistability,ekanem2020signature,kumar2021numerical}, channels with a cylinder \citep{kenney2013large,peng2023numerical} 
or cylinder arrays \citep{khomami1997stability,grilli2013transition,shi2016growth}, porous media \citep{walkama2020disorder,browne2021elastic,haward2021stagnation,kumar2022transport}, and so on.
This apparent gap between abundant experimental and numerical observations and limited theoretical understanding motivates systematic stability analyses of viscoelastic flows with and without inertia in complex geometries.

Such studies remain scarce, most likely owing to the notorious numerical difficulty of simulating viscoelastic flow, namely the so-called high Weissenberg number problem. To the best of our knowledge, and surprisingly, they have been restricted to
global linear stability analyses for
two-dimensional (2D) inertialess flows past a single cylinder~\citep{sureshkumar_linear_1999, smith_finite_2000, spyridakis_viscoelastic_2024} or an array of cylinders~\citep{sureshkumar_linear_1999, smith_finite_2000, smith_linear_2003, sahin2008parallel} in a channel. 
Most of these works used the finite element method (FEM) for spatial discretization, whereas \cite{sahin2008parallel} used a finite volume method.
These works identified distinct instability mechanisms. 
\cite{smith_finite_2000} established that flow of an Oldroyd-B fluid past a single cylinder remains linearly stable for all Weissenberg numbers accessible to their analysis, 
while \cite{spyridakis_viscoelastic_2024} examining a Phan-Thien--Tanner fluid identified strain-rate thinning and elasticity as jointly important for the symmetry-breaking bifurcation. For periodic cylinder arrays, the dominant instability arises from perturbations to the shearing velocity-gradient component interacting with base-flow polymeric stresses~\citep{smith_linear_2003}, or from elastic stresses in inter-cylinder recirculation regions producing spiral vortices \citep{sahin2008parallel}.
In these configurations,  flow instabilities are typically associated with strong shear or curved streamlines, leaving extension-dominated viscoelastic flows in complex geometries largely unexplored.

A representative example is the elastic-instability-induced symmetry breaking of low-Reynolds-number polymeric flows through a nominally symmetric cross-slot microfluidic device. As shown in figure~\ref{fig:sketchAndMesh}(a), it consists of a central square junction connected to four straight side channels arranged orthogonally. Fluid enters through the two opposing horizontal channels and is redirected towards the two vertical outlet branches, producing a stagnation point and 
an extensional flow downstream of that point.
Creeping flow of Newtonian fluids in this device, as expected, preserves the geometric symmetry. By contrast, polymeric flow can break this symmetry through elastic instability: it remains symmetric at low strain rate (low $Wi$), transitions to a steady asymmetric state when $Wi$ exceeds a critical value, and then becomes periodic or even chaotic through a secondary instability as the strain rate, and hence $Wi$, increases further. Notwithstanding being first observed nearly half a century ago~\citep{gardner1982photon} and later rediscovered and attributed to elasticity instability by \citet{arratia2006elastic}, the mechanism by which elasticity breaks symmetry remains elusive and controversial. Possible mechanisms include compressional flow upstream of the stagnation point~\citep{poole2007purely,rocha2009extensibility}, extensional flow downstream of it~\citep{cruz2014new}, geometrical singularity and strong shear associated with  corners~\citep{davoodi2019control,davoodi2021stabilization}, 
and the interplay between fluid elasticity and shear-thinning~\citep{yokokoji2024rheological}.

Using the upper-convected Maxwell model, \citet{poole2007purely} first numerically reproduce steady symmetry breaking in cross-slot viscoelastic flow and attribute the instability to compressional flow upstream of the stagnation point. This view is later indirectly supported by \citet{rocha2009extensibility}, 
who find for finite extendable nonlinear elastic fluids that small corner rounding barely affects the instability onset, thus excluding corner's geometric singularity as the cause.
\cite{cruz2014new} attribute the steady symmetry breaking to the central extensional flow, which strongly stretches the polymers and generates a birefringent strand with substantial normal stresses and stress gradients. 
The innovative optimized-shape cross-slot extensional rheometer (OSCER)~\citep{alves2008design,haward2012optimized} eliminates sharp corners
and enables the generation of a nearly homogeneous extensional flow. Experiments~\citep{haward2013instabilities, haward2016elastic} and simulations~\citep{cruz2018characterization} in this configuration nevertheless still reported steady asymmetric flows via instability, thereby excluding corner singularity or near-corner shear as the essential instability driver.
Combining experiments and simulations, \cite{davoodi2019control} argued that this instability is driven by high streamline curvature and shear rates near corners upon showing the peak values of the Pakdel--McKinley parameter therein. Their placement of a small cylinder at the cross center indicated minimal impact on the instability, thus excluding  the stagnation point as the most critical destabilizing mechanism. The recent work by \citet{yokokoji2024rheological}, combining experiments and simulations, demonstrated that the instability arises from a critical coupling between extensional elasticity at the stagnation point and shear-thinning near boundaries. 
While fluid elasticity triggers the instability, shear-thinning is essential for locking the flow into a steady asymmetric state once the instability occurs. 

While these numerical and experimental studies phenomenologically characterize how instability depends on various aspects of cross-slot flow, they have also led to diverse and sometimes conflicting mechanistic interpretations. A first-principles theoretical prediction of the instability onset and mode structure remains unavailable~\citep{davoodi2021stabilization},  leaving the physical origin of symmetry breaking unresolved nearly half a century after the first experiment~\citep{gardner1982photon}. This difficulty has been recognized as an open problem in non-Newtonian fluid mechanics~\citep{wilson2012open,cruz2014new}. Indeed, even for this 2D inertialess flow, \citet{xi2009mechanism} described linear stability analysis as ``an exceedingly demanding task''.

In this work, we address this long-standing challenge by integrating global linear stability analysis (LSA), structural sensitivity and energy-budget analyses, and direct numerical simulations (DNS) for a creeping flow of Oldroyd-B fluid in a standard 2D cross-slot channel. 
Through these analyses, we identify the physical mechanism underlying the first symmetry-breaking instability of the cross-slot flow.
The paper is organized as follows. We first describe the problem setup and numerical methods in \S\,\ref{sec:ProblemFormulation} and \S\,\ref{sec:numerical}, respectively. After numerically reproducing the classical instability and bifurcation, we present the main results in \S\,\ref{sec:linear}, including the direct eigenmodes from LSA, the adjoint eigenmodes and wavemaker region from structural sensitivity analysis, and the energy budget. Finally, we summarize our findings and discuss their implications in \S\,\ref{sec:conclusion}.

\section{Problem Formulation} \label{sec:ProblemFormulation}
\begin{figure}[!htbp]
	\centering
	\includegraphics[width=1.0\linewidth]{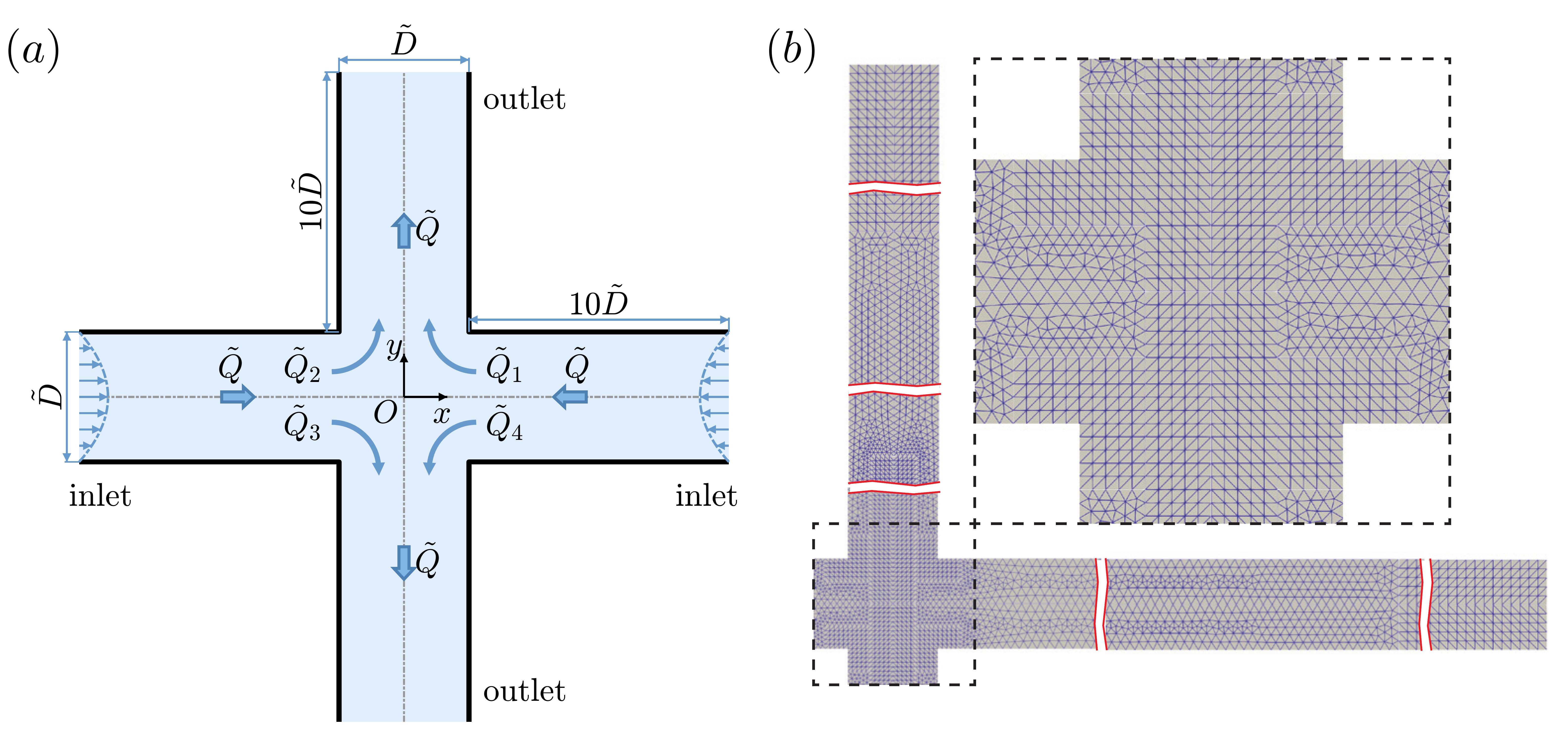}
	\caption{(a) Schematic of the cross-slot geometry and coordinate system. The domain consists of two opposing horizontal inlets and two opposing vertical outlets, each with a width $\tilde{D}$ and length $10\tilde{D}$. The origin $O$ is located at the center, and the flow is divided into four quadrants $(\tilde{Q}_1-\tilde{Q}_4)$. (b) The coarsest mesh, M1, with an inset that highlights the mesh refinement at the central junction.}
	\label{fig:sketchAndMesh}
\end{figure}

We consider a canonical, 2D cross-slot flow of polymeric fluids at vanishing Reynolds number. As shown in figure~\ref{fig:sketchAndMesh}, 
the flow domain comprises a square block of size $\tilde{D}$ connected to four side channels of length $10\tilde{D}$; from hereafter, $\tilde{\left(\bcdot\right)}$ denotes dimensional variables.
Fluid enters the slot through the two horizontal channels, each at an average velocity of $\tilde{U}$, and exits from the two vertical branches. We address the polymeric flow using the incompressible Stokes equations coupled with the diffusive Oldroyd-B constitutive law,
\begin{subequations}
\begin{align}
	& \tnabla \bcdot \tilde{\bsu} = 0, \\
	& -\tnabla \tilde{p} + \mu_s \tnabla \bcdot \left[ \tnabla \tilde{\bsu} + \left(\tnabla \tilde{\bsu}\right)^{\trans} \right] + \tnabla \bcdot \tilde{\bstau} = \bs{0}, 
\end{align}
\end{subequations}
where $\mu_s$ denotes the solvent viscosity out of the total viscosity $\mu$, $\tilde{p}$ and $\tilde{\bsu}$ represent the pressure and fluid velocity, respectively. The polymeric stress tensor  $\tilde{\bstau}$ satisfies
\begin{equation}
	\tilde{\bstau} + \lambda \overset{\triangledown}{\tilde{\bstau}} 
	= \mu_p \left[\tnabla \tilde{\bsu} + \left(\tnabla \tilde{\bsu}\right)^{\trans}\right]
	+ \tilde{\epsilon} \tilde{\nabla}^2 \tilde{\bstau},
\end{equation}
where $\mu_p = \mu -\mu_s$ is the polymer viscosity, $\lambda$ is the relaxation time of polymers, $\tilde{\epsilon}$ is the polymer-stress diffusivity, and $\overset{\triangledown}{\tilde{\bstau}}$ is the upper-convective derivative of $\tilde{\bstau}$,
\begin{equation}
	\overset{\triangledown}{\tilde{\bstau}} 
	= \frac{\partial \tilde{\bstau}}{\partial \tilde{t}} 
    + \tilde{\bsu} 
    \bcdot \tnabla \tilde{\bstau} 
	- \tilde{\bstau} \bcdot \tnabla \tilde{\bsu}
	- (\tnabla \tilde{\bsu})^{\trans} \bcdot \tilde{\bstau},
\end{equation}
where $\tilde{t}$ denotes time.

By choosing $\tilde{D}$, $\tilde{U}$, $\tilde{D}/\tilde{U}$, 
and $\mu \tilde{U}/\tilde{D}$ as the characteristic scales of length, velocity, time, and pressure/stress, respectively, we derive the dimensionless governing equations:
\begin{subequations}\label{eq:gov_non}
\begin{align}
	\bnabla \bcdot \bsu & = 0, \label{eq:mass} \\
	-\bnabla p + \beta \bnabla \bcdot \left[ \bnabla \bsu + \left(\bnabla \bsu\right)^{\trans} \right] + \bnabla \bcdot \bstau  & = \bs{0}, \label{eq:momentum} \\
	\bstau + Wi \left[ \frac{\partial \bstau}{\partial t} + \bsu \bcdot \bnabla \bstau - \bstau \bcdot \bnabla \bsu - \left(\bnabla \bsu\right)^{\trans} \bcdot \bstau \right] & = (1-\beta) \left[ \bnabla \bsu + \left(\bnabla \bsu\right)^{\trans} \right] + \epsilon \nabla^2 \bstau, \label{eq:taup0}    
\end{align}   
\end{subequations}
where $\beta = \mu_s/\mu$ is the ratio of solvent viscosity to the total viscosity, $Wi = \lambda \tilde{U}/\tilde{D}$ is the Weissenberg number, the ratio of the polymer relaxation time to the characteristic flow convection time, and $\epsilon = \tilde{\epsilon}/\tilde{D}^2$ denotes the dimensionless diffusivity of polymeric stress. Throughout this study, we set $\epsilon = 5 \times 10^{-5}$ in the bulk unless otherwise specified.

At both inlets, we impose the analytical velocity and polymeric stress profiles of a fully developed channel flow of Oldroyd-B fluids~\citep{kundu1972small}. 
At the outlets, zero pressure, zero streamwise velocity gradient, and zero transverse velocity component are prescribed. No-slip boundary conditions are applied at all walls. When polymer-stress diffusion is present ($\epsilon >0$) in the bulk, we set $\epsilon = 0$ at the walls following \cite{morozov2022coherent,beneitez2025linear}.
\section{Numerical Methods}\label{sec:numerical}
We solve equation~\eqref{eq:gov_non} numerically using FreeFem++ \citep{hecht2012new}, an open-source FEM library. This choice is motivated by our specific emphasis on stability analysis beyond DNS, together with the successful use of FreeFem++ in a broad range of hydrodynamic stability analyses~\citep{sipp2007global, marquet2008sensitivity,fani2012stability,garnaud2013preferred,lashgari2014planar,boujo2015sensitivity,tammisola2016coherent,fabre2018practical,douglas2026ffbifbox}.

For numerical stabilization, we adopt the discrete elastic-viscous split-stress gradient (DEVSS-G) formulation \citep{liu1998viscoelastic} and the streamline upwind Petrov--Galerkin (SUPG) scheme \citep{baaijens1998mixed} as successfully used in our early studies~\citep{zhu2011locomotion, zhu2012self, su2022viscoelastic, de2026colloidal}, effectively solving the
the following equations:
\begin{subequations} \label{eq:DEVSSG}
\begin{align}
    & \bnabla \bcdot \bsu =0, \\
	& -\bnabla p + \bnabla \bcdot \left[\bnabla \bsu + \left(\bnabla\bsu\right)^{\trans} \right] + \bnabla \bcdot \bstau + \left(\beta-1\right) \bnabla \bcdot \left( \sfbiG+\sfbiG^{\trans} \right) =\boldsymbol{0}, \label{eq:DEVSSG_momentum} \\
	& \bstau+ Wi \left( \frac{\partial \bstau}{\partial t} + \bsu\bcdot\bnabla\bstau - \bstau\bcdot\sfbiG - \sfbiG^{\trans}\bcdot\bstau \right) = \left(1-\beta\right) \left(\sfbiG+\sfbiG^{\trans}\right) + \epsilon \nabla^2 \bstau, \\
    & \sfbiG = \bnabla \bsu,
\end{align}
\end{subequations}
where $\sfbiG$ denotes the continuous approximation of the velocity gradient. The computational domain is discretized with triangular Taylor--Hood elements ($P2$--$P1$) for the velocity--pressure pair and piecewise linear continuous elements ($P1$) for $\bstau$ and $\sfbiG$. 
We solve for $p$, $\bsu$, and $\sfbiG$ in a coupled formulation, while addressing the constitutive equation with the characteristic-Galerkin
method. Further details of the numerical implementation are provided in Appendix \ref{appen:DNS}.

We use three meshes of increasing resolution for the mesh-independence study, which is initiated in the following section \S\,\ref{sec:DNS}
and further detailed in Appendix \ref{appen:mesh}.
The coarsest mesh M1, presented in figure \ref{fig:sketchAndMesh}(b), consists of 13,176 elements and 7,043 nodes, with element sizes ranging from $0.0298$ to $0.1414$. The mesh is locally refined in the central region to better resolve the high polymeric stresses and their sharp gradients expected from the extensional flow therein.
Meshes M2 and M3 are generated by successively refining the central region, reducing the minimum element size by a factor of approximately 2. 

\section{DNS Results on Instability and Bifurcation} \label{sec:DNS}
Upon setting the viscosity ratio to $\beta = 1/9$ throughout this work~\citep{cruz2014new,davoodi2019control,davoodi2021stabilization}, we perform time-dependent DNS to reproduce the classical, steady symmetry-breaking instability of viscoelastic cross-slot flow. It is characterized by a transition from a steady symmetric to a steady asymmetric counterpart with increasing $Wi$. 

\begin{figure}[!htbp]
	\centering
	\includegraphics[width=1\linewidth]{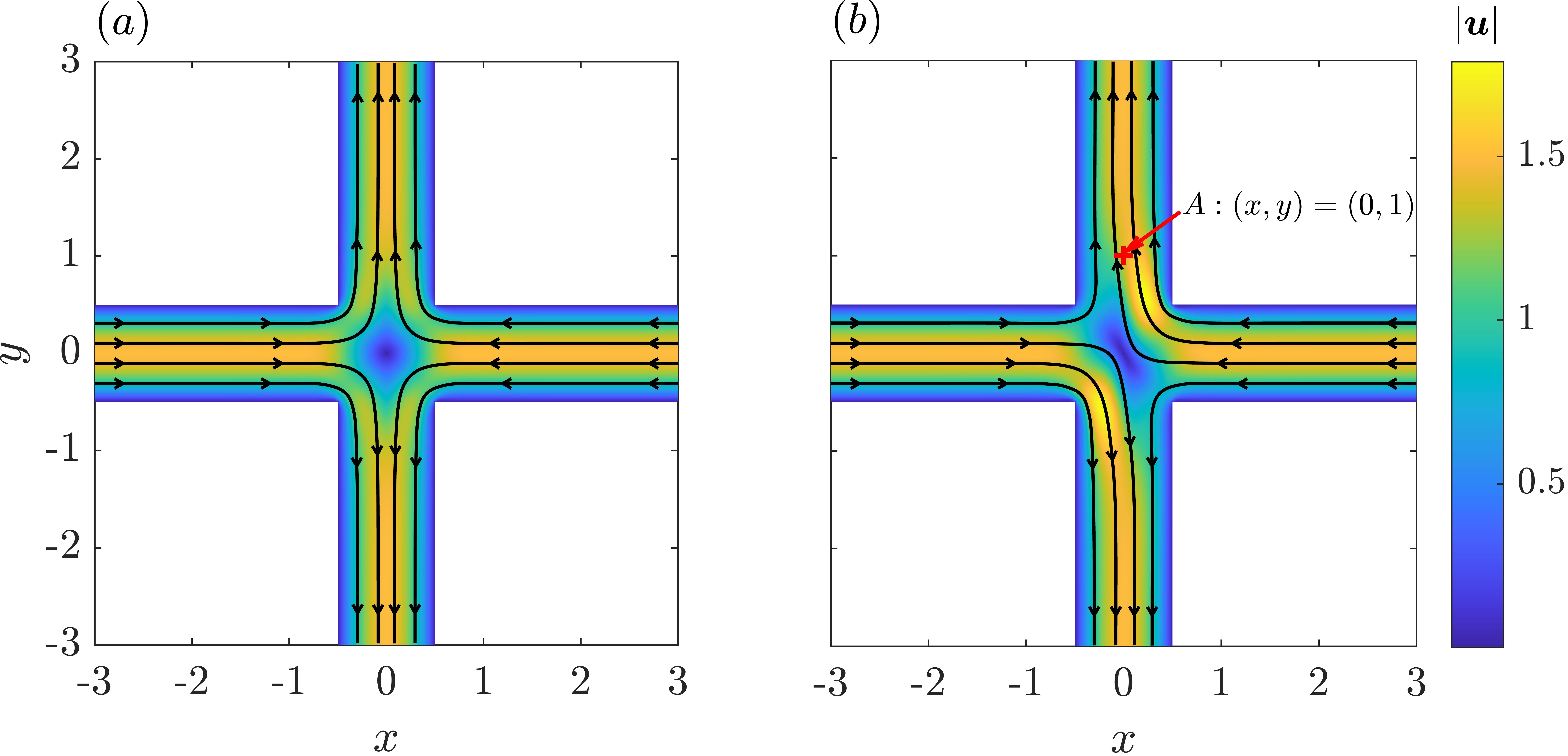}
	\caption{
    Steady flow fields from DNS below and above the symmetry-breaking threshold: (a) stable symmetric state at $Wi=0.3$ and (b) post-bifurcation steady asymmetric state at $Wi=0.38$.
    The fields are visualized by the velocity magnitude $|\bsu|$ with superimposed streamlines.
    A velocity probe is placed at position $A$ to monitor the temporal evolution of the instability in \S\,\ref{fig:MN40_fitSigmaLandau_eps5e-5_1r2c}.
}
	\label{fig:MN40_DNS_Wi0.3_0.38}
\end{figure}

Figure~\ref{fig:MN40_DNS_Wi0.3_0.38} shows the numerical velocity fields of the stable symmetric state at $Wi=0.3$ and the post-bifurcation asymmetric state at $Wi=0.38$, obtained after the simulation reaches a steady regime. At $Wi=0.3$, the flow possesses reflection-symmetry about both the horizontal and vertical axes and centrosymmetry about the origin. At $Wi=0.38$, the flow presents a skewed pattern, losing both reflection symmetries while retaining centrosymmetry. As documented previously~\citep{arratia2006elastic,poole2007purely}, 
this flow asymmetry in the cross-slot channel results in an imbalanced distribution of flux within the central junction. Following \citet{rocha2009extensibility,cruz2014new}, we quantify this asymmetry using the normalized flux imbalance,
\begin{equation}
	\Delta Q = \frac{\left(Q_2+Q_4\right)-\left(Q_1+Q_3\right)}{Q_1+Q_2+Q_3+Q_4},
\end{equation}
where $Q_1$, $Q_2$, $Q_3$, and $Q_4$ denote the partial fluxes through the four diagonal cross-sections connecting the center to the corners [see figure \ref{fig:sketchAndMesh}(a)]. By definition, $\Delta Q=0$ for a symmetric flow given $Q_1=Q_2=Q_3=Q_4$.

\begin{figure}[!htbp]
	\centering
	\includegraphics[width=1.0\linewidth]{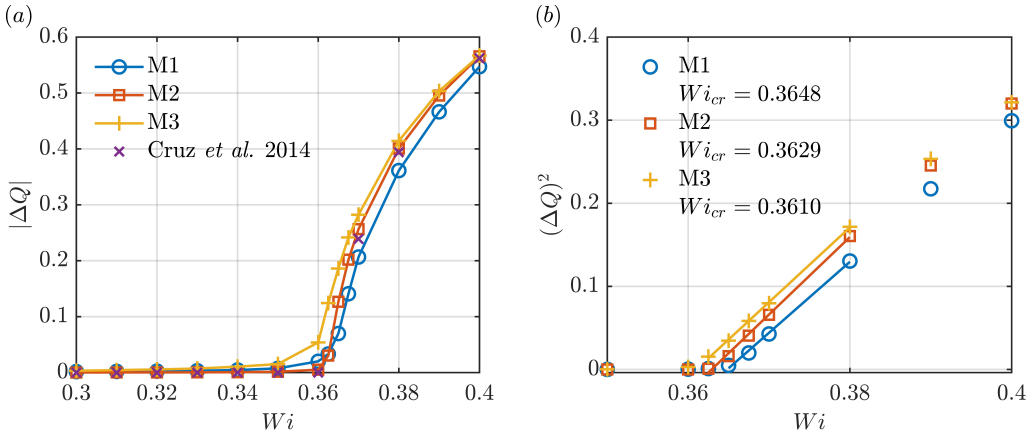}
	\caption{
        (a) Absolute flux imbalance $|\Delta Q|$ versus $Wi$ on meshes M1, M2, and M3, compared with the numerical data of \citet{cruz2014new}.
		(b) $\left( \Delta Q\right)^2$ versus $Wi$; linear extrapolation to zero yields the critical Weissenberg number $Wi_{cr} = 0.3648$, $0.3629$, and $0.3610$ for M1, M2, and M3, respectively.}
	\label{fig:DQ_eps5e-5_subfig}
\end{figure}

We present in figure~\ref{fig:DQ_eps5e-5_subfig} the absolute flux imbalance $|\Delta Q|$ (a) and its square $\left( \Delta Q \right)^2$ (b) as functions of $Wi$ for all three meshes, M1, M2, and M3. 
The data exhibit a consistent qualitative trend: $\Delta Q=0$ remains zero until the onset of instability at a critical Weissenberg number $Wi_{cr}\approx 0.36$. Above this threshold, $\left( \Delta Q \right)^2$ increases linearly with $Wi-Wi_{cr}$ [see figure~\ref{fig:DQ_eps5e-5_subfig}(b)], characteristic of the well-recognized supercritical pitchfork bifurcation. Extrapolating the linear fits to $\left( \Delta Q \right)^2=0$ yields the thresholds, $Wi_{cr}=0.3648$, $0.3629$, and $0.3610$ for M1, M2, and M3, respectively, indicating only weak mesh dependence of $Wi_{cr}$. 
However, an anomalous mesh dependence of $|\Delta Q|$ is observed near $Wi_{cr}$, which we infer is due to the finite polymer-stress diffusivity $\epsilon$, as discussed in Appendix \ref{appen:mesh}.
Unless otherwise specified, we use mesh M2 in the remainder of the work.

\section{Global Linear Stability, Structural Sensitivity, and Energy--Budget Analyses}\label{sec:linear}

\subsection{Global linear stability analysis}
 \label{subsec:lsa}
Having revealed the DNS results, we conduct global LSA  to better understand the symmetry-breaking instability.
As commonly done, we decompose the state variable $\bsq(\bs{x},t) = [p,\bsu, \sfbiG, \bstau](\bs{x},t)$ into a steady base state $\bsq^b(\bs{x})$ and an infinitesimal perturbation $\bsq' (\bs{x},t)$, 
\begin{equation} \nonumber
	\bsq(\bs{x},t) = \bsq^b(\bs{x}) + \bsq'(\bs{x},t),
\end{equation}
where $\bs{x}$ is the spatial position vector and $\bsq^b(\bs{x})$ is obtained by solving the steady-state governing equations via the Newton--Raphson method, see Appendix \ref{appen:LSA} for more details. 
We linearize the DEVSS-G system~\eqref{eq:DEVSSG} about the base state and seek perturbations in the normal-mode form
\begin{equation} \label{eq:normal_mode_ansatz}
    \bsq'(\bs{x},t) = e^{\sigma t} \hat{\bsq}(\bs{x}),
\end{equation}
where $\sigma $ 
is the complex eigenvalue, $\hat{\bsq}(\bs{x})=[\hat{p},\hat{\bsu}, \hat{\sfbiG}, \hat{\bstau}](\bs{x})$ is the corresponding eigenmode, and $\Real(\sigma)$ and $\Imag(\sigma)$ denote its temporal growth rate and oscillation frequency, respectively. From hereafter, we refer to $\hat{\bsq}$ as the direct eigenmode, in contrast to the adjoint eigenmode introduced later in \S\,\ref{sec:sensitivity}.   Substituting \eqref{eq:normal_mode_ansatz} into the linearized perturbation equations yields the direct global stability eigenvalue problem
\begin{subequations} \label{eq:lsa}
\begin{align}
	& \bnabla \bcdot \hat{\bsu} =0, \label{eq:lsaMass} \\  
	& -\bnabla \hat{p} + \bnabla \bcdot \left[ \bnabla \hat{\bsu} + \left(\bnabla \hat{\bsu}\right)^{\trans} \right] + (\beta-1) \bnabla \bcdot \left( \hat{\sfbiG} + \hat{\sfbiG}^{\trans} \right) +\bnabla \bcdot \hat{\bstau} =\bs{0}, \label{eq:lsaMomentum} \\
	& \hat{\bstau} + Wi \overset{\square}{\hat{\bstau}} 
		= \left(1-\beta\right) \left(\hat{\sfbiG} + \hat{\sfbiG}^{\trans}\right)
		+ \epsilon \nabla^2 \hat{\bstau}, \label{eq:lsaTau} \\
    & \hat{\sfbiG} = \bnabla \hat{\bsu},
\end{align}
\end{subequations}
where $\overset{\square}{\hat{\bstau}}$ is the linearized upper-convective derivative of $\hat{\bstau}$,
\begin{equation} \label{eq:upcTau}
	\overset{\square}{\hat{\bstau}} 
	= \sigma \hat{\bstau} 
	+ \bsu^{b} \bcdot \bnabla \hat{\bstau}
	+ \hat{\bsu} \bcdot \bnabla \bstau^{b}
	-\bstau^{b} \bcdot \hat{\sfbiG}
	-\hat{\bstau} \bcdot \sfbiG^{b}
	-\left(\sfbiG^{b}\right)^{\trans} \bcdot \hat{\bstau}
	-\hat{\sfbiG}^{\trans} \bcdot \bstau^{b}. 
\end{equation}
The system \eqref{eq:lsa} is subject to the following boundary conditions: zero velocity and stress perturbations at the inlets, vanishing pressure and transverse velocity perturbations at the outlets, and zero velocity perturbation at all 
walls. Derivation of the eigenvalue problem \eqref{eq:lsa} and its numerical implementation are provided in Appendix \ref{appen:LSA}.

\begin{figure}[!htbp]
	\centering
	\includegraphics[width=1.0\linewidth]{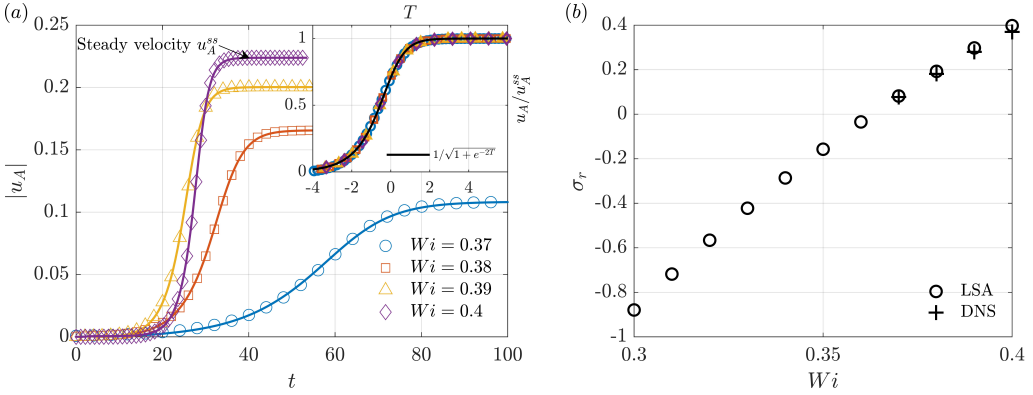}
	\caption{(a) Time evolution of $|u_A|$ at point $A: (x,y) = (0,1)$ for $Wi = 0.37$, $0.38$, $0.39$, and $0.40$; curves show Landau--equation fits. Inset: rescaled velocity $u/u_A^{ss}$ as a function of the rescaled time $T = \sigma_r(t - t_h)$; data for all $Wi$ collapse onto the Landau solution $1/\sqrt{1+e^{-2T}}$ (solid curve). (b) Leading temporal growth rate $\sigma_r$ versus $Wi$: LSA results (circles) compared with the DNS counterparts (crosses). }
	\label{fig:MN40_fitSigmaLandau_eps5e-5_1r2c}
\end{figure}

Using FreeFem++, we solve the discretized form of equation~\eqref{eq:lsa} for the complex eigenvalues $\sigma$ and the corresponding eigenmodes. Each eigenmode is normalized such that the associated  eigenvector of the generalized matrix eigenvalue problem~\eqref{Eq:GEignSys} has unit $\ell^2$-norm.
We then assess the LSA prediction by comparing the real part $\sigma_r = \Real(\sigma)$ of  
the leading eigenvalue $\sigma$ with the temporal growth rate extracted from DNS. 
To achieve so, we monitor the time evolution of the $x$-velocity-component, $u_A$, at a probe 
$A: (x,y) = (0,1)$ placed downstream of the stagnation point [see figure~\ref{fig:MN40_DNS_Wi0.3_0.38}(b)]; the magnitude $|u_A|$ exhibits an initial exponential growth before saturating at a steady value $|u^{ss}_A|$. 
Following the Landau amplitude equation~\citep{landau1944problem,stuart1960non,drazin2004hydrodynamic}, we fit the DNS data through
\begin{equation}
    u_A(t) = \frac{u_A^{ss}}{\sqrt{1 + e^{-2\sigma_r(t - t_h)}}},
\end{equation}
treating $\sigma_r$ and the half-saturation time $t_h$ when $u_A = u_A^{ss}/\sqrt{2}$ as fitting parameters. The temporal signals $|u_A|(t)$ and their corresponding fitting curves for four unstable $Wi$ values are depicted in figure~\ref{fig:MN40_fitSigmaLandau_eps5e-5_1r2c}(a). Its inset further indicates that the data collapse under the rescaling
\begin{equation}
    \frac{u_A(T)}{u_A^{ss}} = \frac{1}{\sqrt{1 + e^{-2T}}},
\end{equation}
with $T = \sigma_r(t - t_h)$,
supporting a Landau-type amplitude description of the instability growth and saturation.

Figure~\ref{fig:MN40_fitSigmaLandau_eps5e-5_1r2c}(b) illustrates the monotonic increase of the growth rate $\sigma_r$ with $Wi$, with $\sigma_r$ crossing zero at the critical value $Wi_{cr}$ that marks the onset of instability. For $Wi>Wi_{cr}$, the LSA predictions agree reasonably well with the DNS-derived growth rates, validating our implementation of the eigenvalue problem~\eqref{eq:lsa}. For $Wi<Wi_{cr}$, precise extraction of the decay rate from DNS is difficult, and these values are therefore omitted; this omission does not affect the validation of the LSA implementation. In parallel, the LSA-predicted imaginary part $\sigma_i=\Imag(\sigma)$ of the leading eigenvalue $\sigma$ remains negligible across the investigated range of $Wi$, consistent with the steady symmetry-breaking character of the pitchfork bifurcation.

\begin{figure}[!htbp]
	\centering
	\includegraphics[width=1\linewidth]{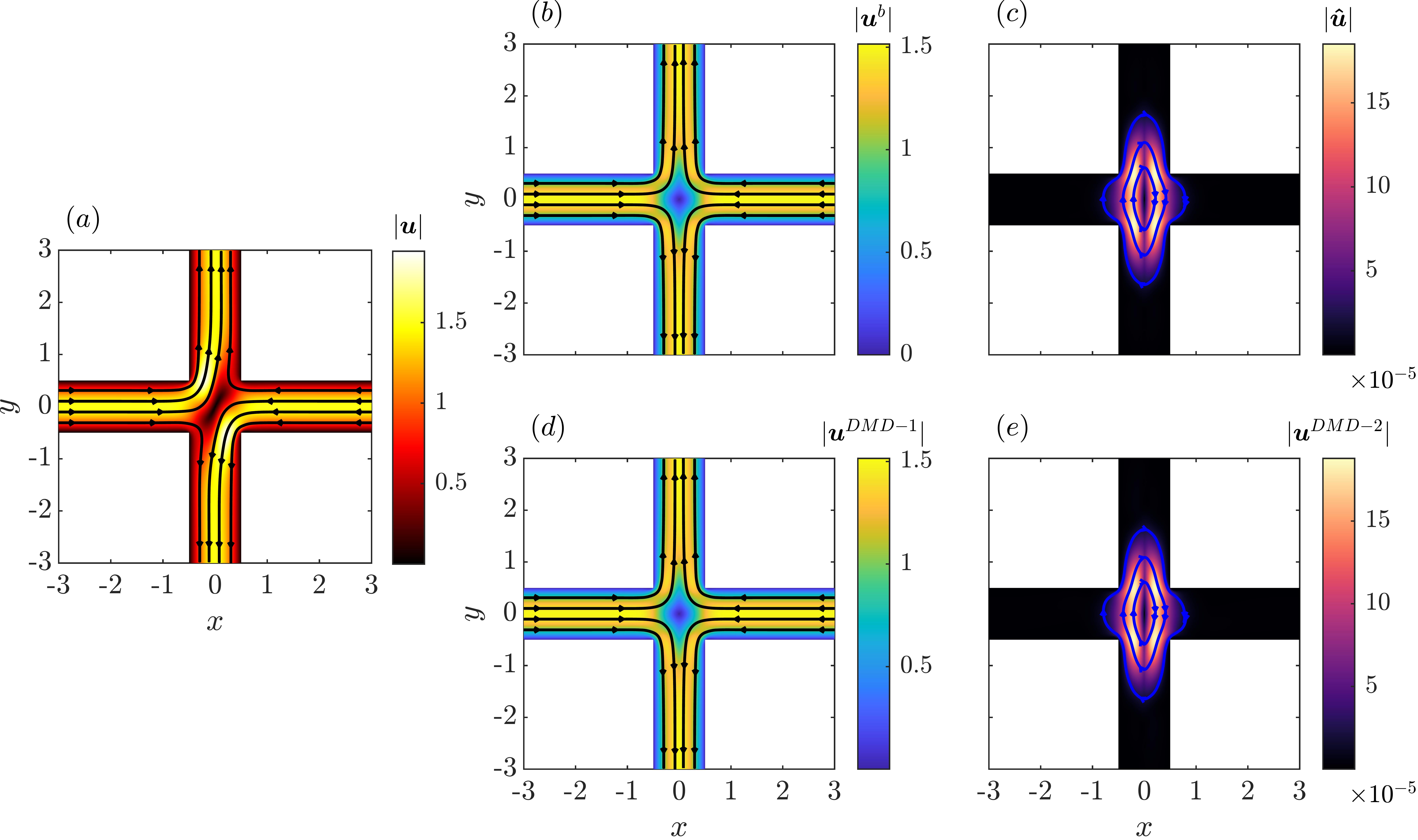}
	\caption{Velocity magnitudes with streamlines at $Wi=0.4$.
		(a) DNS steady asymmetric state $|\bsu|$,
		(b) base flow $|\bsu^{b}|$,
		(c) leading direct eigenmode $|\hu|$,
		(d) and (e) the leading and second DMD modes, $|\bsu^{DMD-1}|$ and $|\bsu^{DMD-2}|$, respectively.
        }
	\label{fig:MN40_u_dns_bas_lsa_Wi0.4}
\end{figure}

Having verified the growth rate, we now examine the velocity field at $Wi=0.4$,
starting from the steady-state asymmetric flow field $\bs{u}$ from DNS as displayed in figure~\ref{fig:MN40_u_dns_bas_lsa_Wi0.4}(a). 
By symmetry, an equivalent steady state can be obtained by reflecting this field about either the $x$- or $y$-axis, see figure~\ref{fig:MN40_DNS_Wi0.3_0.38}(b) for example.
Next, the base flow $\bsu^b$ and leading direct eigenmode $\hat{\bsu}$ are shown in figure~\ref{fig:MN40_u_dns_bas_lsa_Wi0.4}(b) and (c), respectively. 
Beyond DNS and LSA, we further apply dynamic mode decomposition (DMD) (see Appendix \ref{appen:dmd}) to DNS snapshots from the linear growth regime for a data-driven estimate of the leading continuous-time eigenvalues and associated modes~\citep{schmid2010dynamic,rowley2009spectral}. The dominant DMD mode has $\omega^{DMD-1}\approx 0$ and closely reproduces the steady base state [figure~\ref{fig:MN40_u_dns_bas_lsa_Wi0.4}(b, d)]. This state
stands for the mean flow, exhibiting reflection symmetry across both the horizontal and vertical axes with a central stagnation point and an associated extensional flow. 

The next dynamically relevant mode has $\Real(\omega^{DMD-2})\approx 0.372$ and $\Imag(\omega^{DMD-2})\approx0$, consistent with the stationary nature of the instability. Its growth rate agrees closely the  DNS estimate $\sigma_r\approx0.370$ and reasonably with the LSA prediction of $0.398$. Notably, its spatial structure matches the leading direct eigenmode from LSA [figure~\ref{fig:MN40_u_dns_bas_lsa_Wi0.4}(c, e)], thereby cross-validating the DNS, LSA, and DMD results. 
This velocity eigenmode forms a diamond-shaped vortical structure with closed streamlines centered at the stagnation point, and its components exhibit the component-wise parity expected for a centered vortex. Such a mode represents a linear signature of the symmetry-breaking mechanism of the viscoelastic cross-slot flow, whereby a chiral velocity perturbation around the stagnation point deflects the separatrices and produces unequal flow partitioning within the two vertical branches.

\begin{figure}[!htbp]
    \centering
    \includegraphics[width=1\linewidth]{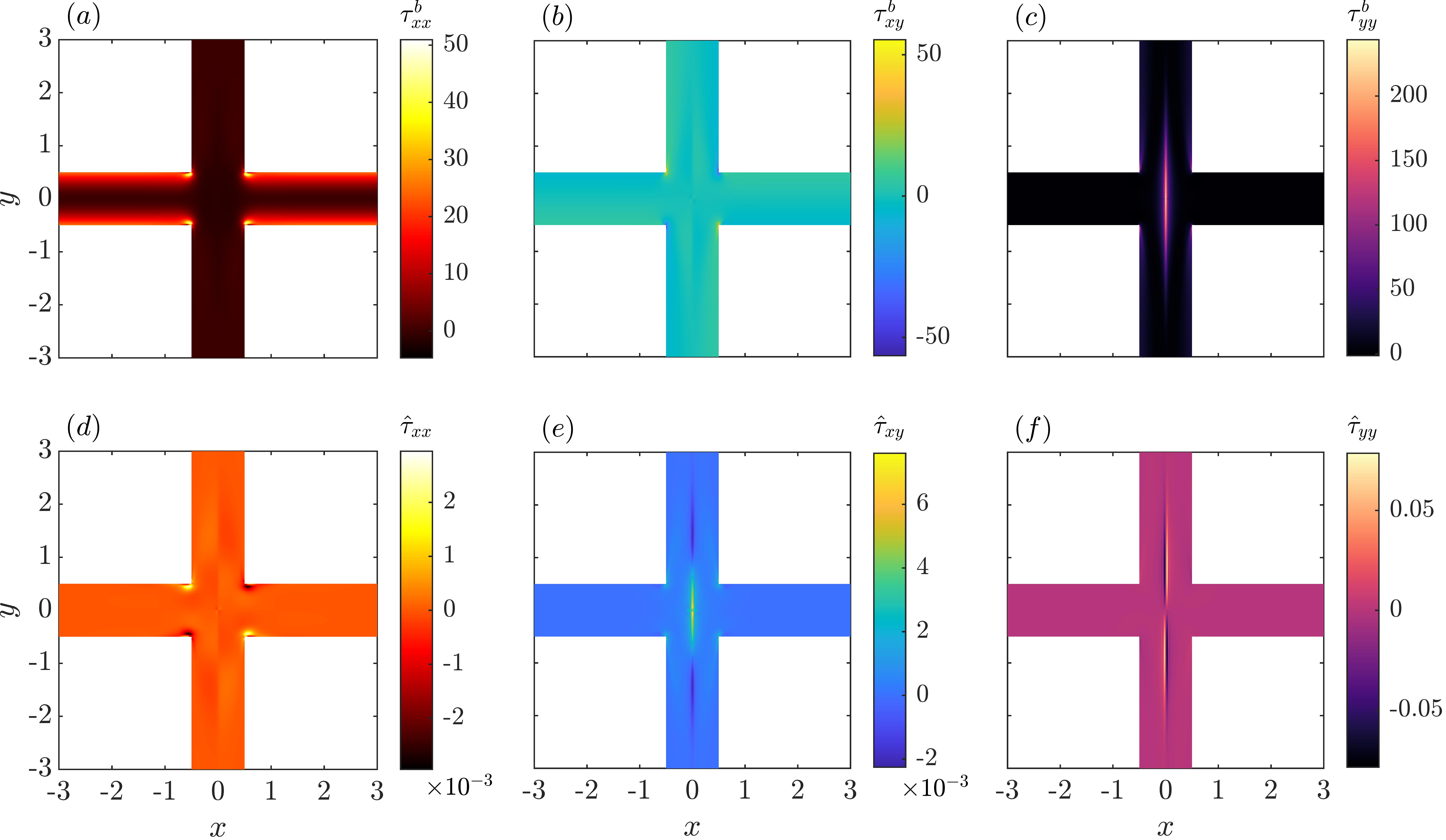}
    \caption{   
    Polymeric stress components at $Wi=0.4$. Top row (a), (b), (c): base state $\tau^{b}_{xx}$, $\tau^{b}_{xy}$, $\tau^{b}_{yy}$. Bottom row (d), (e), (f): leading direct eigenmode $\hat{\tau}_{xx}$, $\hat{\tau}_{xy}$, $\hat{\tau}_{yy}$.}
    \label{fig:MN40_tau_BaseLSA_Wi0.4}
\end{figure}

We then examine the polymeric stress distribution in figure~\ref{fig:MN40_tau_BaseLSA_Wi0.4}, specifically its base state $\bstau^b$ (top row) and leading direct eigenmode $\hat{\bstau}$ (bottom row).
In the base state, $\tau_{xx}^{b}$ and $\tau_{yy}^{b}$ exhibit double reflection symmetry---being symmetric about both the horizontal and vertical axes---whereas $\tau_{xy}^{b}$ exhibits quadrupolar symmetry, being antisymmetric about each axis and invariant under a $180^{\circ}$ rotation about the center.
Notably, $\tau_{yy}^{b}$ is concentrated along a prominent vertical filament through the cross center---a birefringent strand associated with strong polymer extension and alignment in the extensional flow downstream of the stagnation point.
In contrast, the leading eigenmode displays the opposite parity: 
$\hat{\tau}_{xx}$ and $\hat{\tau}_{yy}$ exhibit quadrupolar symmetry, whereas $\hat{\tau}_{xy}$ displays double reflection symmetry. 
This parity reversal suggests that the instability is driven by
the antisymmetric normal-stress perturbation $\hat{\tau}_{yy}$ that tilts and rotates the birefringent strand. 
DNS corroborates this mechanism by capturing the clockwise rotation of the strand 
(figure~\ref{fig:MN40_tau_DNS_Wi0.4}). 
Considering the vortical velocity perturbation shown in figure~\ref{fig:MN40_u_dns_bas_lsa_Wi0.4},  the polymer-stress disturbance substantiates the linear signature of the spontaneous symmetry-breaking in the cross-slot flow. 
Expectedly, the DMD polymeric-stress modes (figure~\ref{fig:MN40_tau_DMD_mode12_Wi0.4}) mirror the velocity modes: the dominant and second modes recover the base state and the leading direct eigenmode, respectively.

\begin{figure}[!htbp]
	\centering
	\includegraphics[width=1\linewidth]{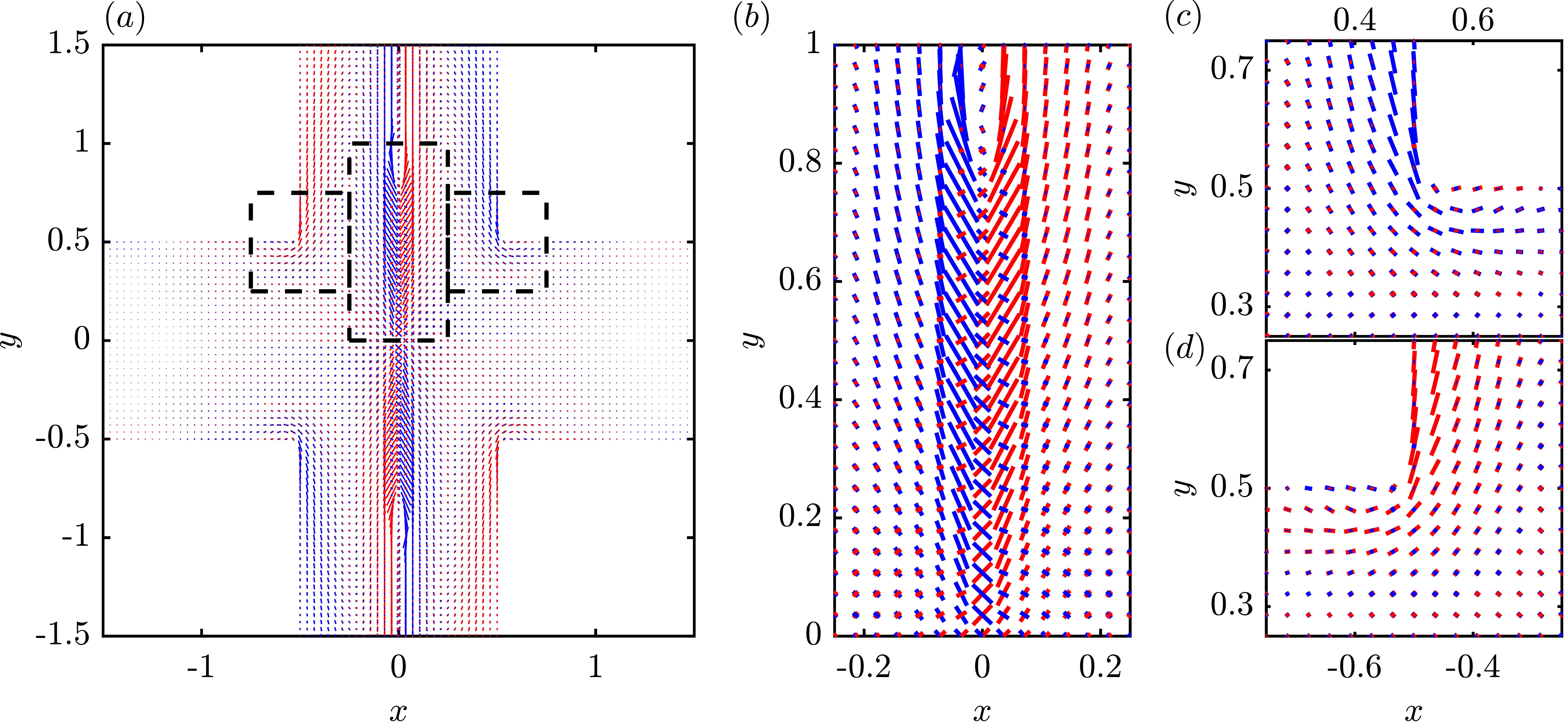}
	\caption{A pair of color-coded (red or blue) segments at each position indicates principal directions, $\bs{p}_1$ and $\bs{p}_2$, of the local polymer-conformation perturbation $\hat{\mathsfbi{C}}$ when $Wi=0.4$. The segment lengths are scaled by $|\delta\lambda_{1,2}|^{1/2}$. Red and blue segments denote positive and negative $\delta\lambda_{i}$, respectively, corresponding to stretching and compression perturbations.
    }
	\label{fig:sticks_LSA_Wi0.4}
\end{figure}

To seek a more intuitional  understanding of the polymer microstructural disturbance, we examine the perturbation conformation tensor
\begin{equation} \nonumber
    \hat{\mathsfbi{C}} = \frac{Wi}{1-\beta} \hat{\bstau},
\end{equation} 
which represents the eigenmode of the polymer conformation perturbation corresponding to the leading polymeric-stress eigenmode. It is real and symmetric, therefore admitting the spectral decomposition
\begin{equation} \label{eq:Chat_decomp}
	\hat{\mathsfbi{C}} = \delta\lambda_1 \,\bs{p}_1\bs{p}_1 + \delta\lambda_2 \,\bs{p}_2 \bs{p}_2,
\end{equation}
where $\bs{p}_1$ and $\bs{p}_2$ denote the orthonormal principal directions of the local polymer-conformation perturbation, and $\delta\lambda_i$ ($|\delta\lambda_1| \geq |\delta\lambda_2|$) are the corresponding principal values. Positive and negative $\delta\lambda_i$ correspond to stretching and compression perturbations along $\bs{p}_i$, respectively. 
Inspired by the visualization of \citet{Teran2010viscoelastic},  
figure~\ref{fig:sticks_LSA_Wi0.4} depicts 
the spatial distribution of $\bs{p}_{1,2}$ in the form of two orthogonal segments with lengths proportional to $|\delta\lambda_{1,2}|^{1/2}$. The segments are color-coded by the sign of $\delta \lambda_{1,2}$---red for positive (stretching) and blue for negative (compression), respectively.

This perturbation is weak near the corners [figure~\ref{fig:sticks_LSA_Wi0.4}(c)] and does not peak right at the stagnation point. 
Instead, it is most pronounced along two vertical bands above and below the stagnation point, each resembling a herringbone structure. As shown in figure~\ref{fig:sticks_LSA_Wi0.4}(b) for the upper ``herringbone'', 
its left-branch oblique ``ribs'' represent polymeric compression along an approximate backslash direction, whereas the right-hand counterparts highlight polymeric stretching in an approximate slash direction.  This simultaneous side-by-side stretching and compression naturally induces a rotational reorientation of the polymer microstructure along the vertical centerline, thereby triggering the elastic instability. This interpretation is consistent with the rotational perturbation velocity $\hu$ [figure~\ref{fig:MN40_u_dns_bas_lsa_Wi0.4}(c)] and the antisymmetric normal-stress perturbation  $\hat{\tau}_{yy}$ [figure~\ref{fig:MN40_tau_BaseLSA_Wi0.4}(f)].

\subsection{Structural sensitivity analysis} \label{sec:sensitivity}
While the eigenmodes reveal where the instability is mostly active and hence provide useful insight into the instability mechanism,
they cannot pinpoint which regions of the flow are most receptive to external forcing or structural modifications. This complementary information is inherently encoded in the adjoint eigenmode $\hat{\bsq}^{\dagger}(\bs{x}) = \left[ \hat{p}^{\dagger}, \hat{\bsu}^{\dagger}, \hat{\sfbiG}^{\dagger}, \hat{\bstau}^{\dagger}\right](\bs{x})$.
As detailed in Appendix \ref{appen:adjoint}, we derive the adjoint linearized equation \eqref{eq:adjointG_apd} for the adjoint state variable $\bsq^\dagger(\bs{x},t) = \left[ p^{\dagger}, \bsu^{\dagger}, \mathsfbi{G}^{\dagger}, \bstau^{\dagger}\right](\bs{x},t)$ by adapting the framework of \citet{kim2023adjoint}, originally developed for drag-sensitivity analysis in viscoelastic flow past a cylinder.
Substituting the normal-mode form $\bsq^{\dagger}(\bs{x},t) = e^{-\sigma^{\dagger} t}\hat{\bsq}^{\dagger}(\bs{x})$ into equation \eqref{eq:adjointG_apd} yields the adjoint eigenvalue problem:
\begin{subequations}\label{eq:adj_eigen}
\begin{align}
&\bnabla \bcdot \hat{\bsu}^{\dagger} = 0, \\
& \bnabla \cdot \left[ \bnabla \hat{\bsu}^{\dagger} + \left(\bnabla \hat{\bsu}^{\dagger}\right)^{\trans} \right]
+ (\beta-1)\bnabla \bcdot \left[\hat{\sfbiG}^{\dagger}+\left(\hat{\sfbiG}^{\dagger}\right)^{\trans}\right] - \bnabla \hat{p}^{\dagger} 
+ Wi \bnabla\bstau^{b} \bs{:} \hat{\bstau}^{\dagger} \nonumber \\
& \quad + 2Wi\bnabla \bcdot \left(\bstau^{b} \bcdot \hat{\bstau}^{\dagger}\right) 
+ 2\left(1-\beta\right) \bnabla \bcdot \hat{\bstau}^{\dagger} = \bs{0},  \\
&\hat{\bstau}^{\dagger} + Wi\left[ \sigma^{\dagger} \hat{\bstau}^{\dagger} - \bsu^{b} \bcdot \bnabla\hat{\bstau}^{\dagger} - \bnabla \bsu^{b} \bcdot \hat{\bstau}^{\dagger} 
- \hat{\bstau}^{\dagger} \bcdot \left(\bnabla \bsu^{b}\right)^{\trans} \right] 
- \frac{1}{2}\left[\hat{\sfbiG}^{\dagger} + \left(\hat{\sfbiG}^{\dagger}\right)^{\trans}\right]
- \epsilon \nabla^2 \hat{\bstau}^{\dagger} = \bs{0}, \\
& \hat{\sfbiG}^{\dagger} = \bnabla \hat{\bsu}^{\dagger}.
\end{align} 
\end{subequations}

\begin{figure}[!htbp]
    \centering
    \includegraphics[width=1\linewidth]{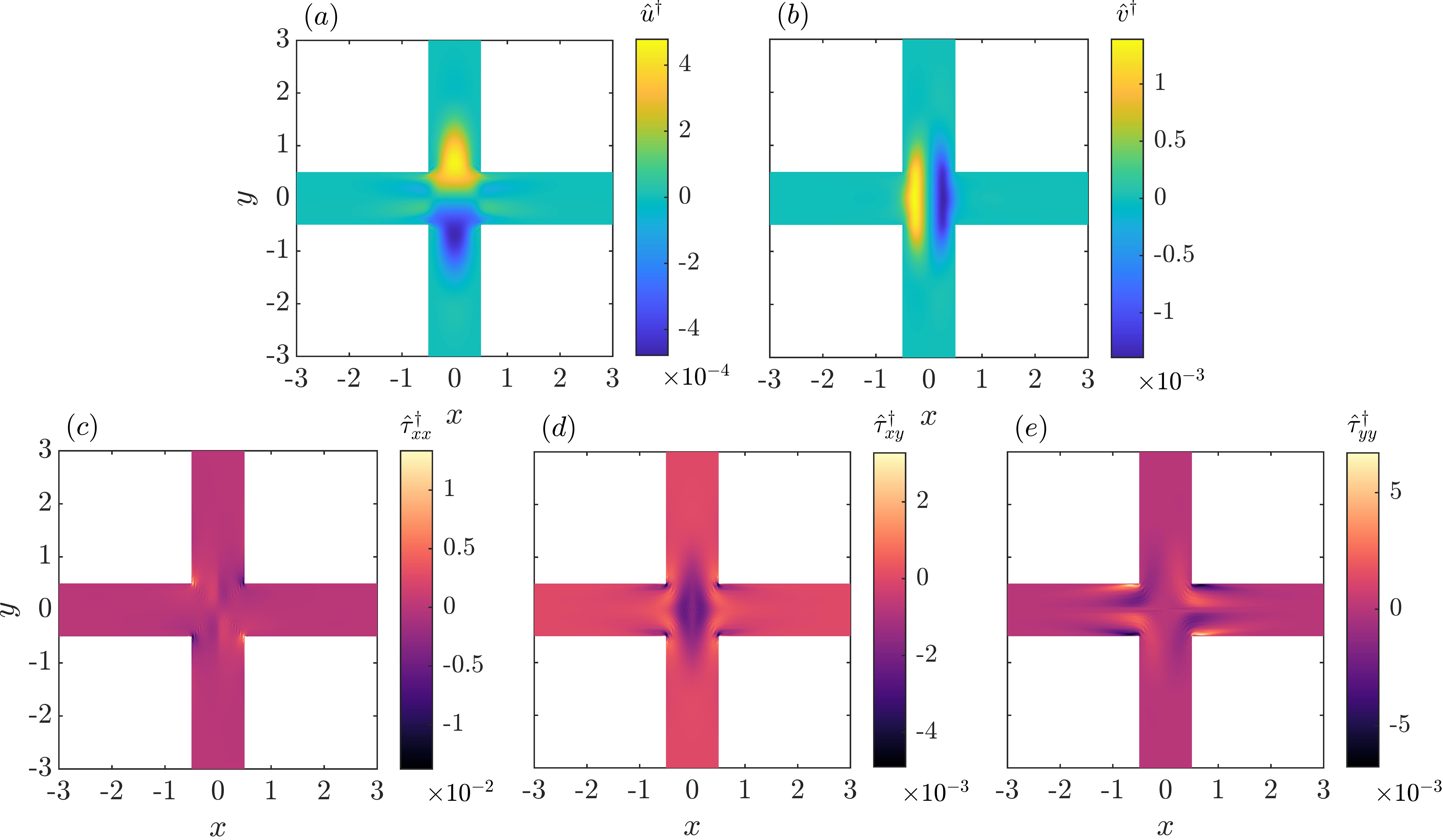}
    \caption{Adjoint eigenmode at $Wi=0.4$ including its
    (a--b) velocity components $\hat{u}^\dagger$ and $\hat{v}^\dagger$, and 
    (c--e) polymeric-stress components $\hat{\tau}_{xx}^\dagger$, $\hat{\tau}_{xy}^\dagger$, and $\hat{\tau}_{yy}^\dagger$.}
    \label{fig:MN40_adj_Wi0.4}
\end{figure}

Equation \eqref{eq:adj_eigen} is solved similarly as the eigenvalue problem. The velocity and stress components of the adjoint eigenmode at $Wi = 0.4$ are presented in figure~\ref{fig:MN40_adj_Wi0.4}, encoding the spatial distribution of the receptivity of the instability to external forcing. 
The adjoint velocity components are localized in the central junction region, with $\hat{u}^\dagger$ and $\hat{v}^\dagger$ being antisymmetric about the horizontal ($y=0$) and vertical ($x=0$) axes, respectively. This spatial distribution indicates that the instability is most receptive to horizontal momentum perturbations downstream of the stagnation point and vertical momentum perturbations upstream of it.
For the adjoint stress components, $\hat{\tau}_{xx}^\dagger$ is concentrated at the 
corners with a sign pattern that is antisymmetric about the geometric diagonal, mirroring the structure of the direct eigenmode $\hat{\tau}_{xx}$. 
The shear component $\hat{\tau}_{xy}^\dagger$ fills both inflow arms with positive values while exhibiting a negative region at the central stagnation point and near the corners.
$\hat{\tau}_{yy}^\dagger$ is concentrated near the 
corners, with an antisymmetric pattern analogous to $\hat{\tau}_{xx}^\dagger$.

The direct and adjoint eigenmodes exhibit markedly different spatial structures, so either mode alone provides limited insight into the instability mechanism. Following \citet{giannetti2007structural,pralits2010instability}, we combine the two eigenmodes to define the velocity-based structural sensitivity field,
\begin{align}\label{eq:eta_u}
    \eta_{\bsu}(\bs{x}) =
    \frac{\left|\hat{\bsu}(\bs{x})\right|
    \left|\hat{\bsu}^{\dagger}(\bs{x})\right|}
    {\left|\int_{\Omega}
    \hat{\bsu} \bcdot \hat{\bsu}^{\dagger *}\,\dd\Omega\right|},
\end{align}
which measures the spatial overlap between the direct velocity eigenmode $\hat{\bsu}$ and the adjoint velocity eigenmode $\hat{\bsu}^{\dagger}$. Here, $\Omega$ denotes the computational domain, and $*$ denotes complex conjugation. The classical structural sensitivity analysis, originally developed for inertia-induced instabilities in Newtonian flows, identifies the region where the leading eigenvalue is most sensitive to localized velocity feedback---equivalently---where both the direct and adjoint velocity eigenmodes are large. This region is referred to as the wavemaker often interpreted as the core or origin of instability~\citep{giannetti2007structural,luchini2014adjoint,schmid2014analysis}, since it marks where the global mode has both strong local amplitude and high receptivity to feedback.
For the viscoelastic flow instability considered here, the relevant perturbation dynamics involve not only velocity perturbations but also polymeric stress perturbations. Hence, we introduce an analogous stress structural sensitivity field,
\begin{align}\label{eq:eta_tau}
    \eta_{\bstau}(\bs{x}) =
    \frac{\left|\hat{\bstau}(\bs{x})\right|
    \left|\hat{\bstau}^{\dagger}(\bs{x})\right|}
    {\left|\int_{\Omega}
    \hat{\bstau} \bs{:} \hat{\bstau}^{\dagger *}\,\dd\Omega\right|}.
\end{align}

\begin{figure}[!htbp]
    \centering
    \includegraphics[width=1\linewidth]{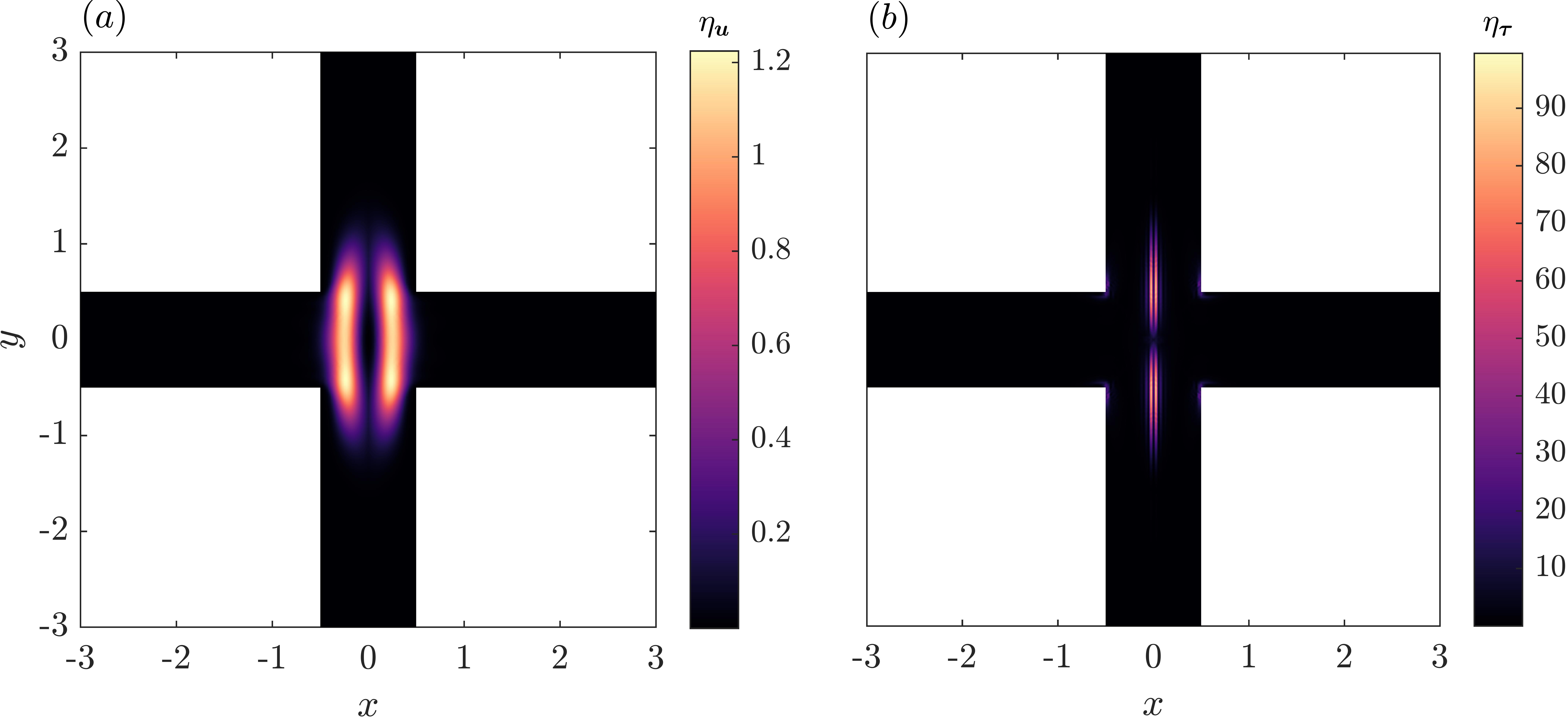}
    \caption{
     Velocity-based (a) and stress-based (b) structural sensitivity fields, $\eta_{\bsu}$ (equation~\ref{eq:eta_u}) and  $\eta_{\bstau}$ (equation~\ref{eq:eta_tau}),  when $Wi=0.4$.
    }
    \label{fig:MN40_wavemaker_Wi0.4}
\end{figure}

As shown in figure~\ref{fig:MN40_wavemaker_Wi0.4} for $Wi=0.4$, the velocity-based sensitivity field $\eta_{\bsu}$ attains large values in two symmetric lobes straddling the $y$-axis, identifying the velocity-based wavemaker region of the instability. Compared to $\eta_{\bsu}$,  the stress-based sensitivity field $\eta_{\bstau}$ is much more strongly localized, indicating that polymeric-stress perturbations play a central role in the elastic instability. 
The stress-based wavemaker is concentrated along the central, high-strain-rate stripe of the extensional flow: 
it consists of a pair of vertical filaments located slightly above the stagnation point, together with a reflection-symmetric pair below it. This structure closely resembles the herringbone-like pattern of the concentrated polymer-conformation perturbation shown in figure~\ref{fig:sticks_LSA_Wi0.4}, suggesting that the stress-based wavemaker captures the core of the elastic instability. By contrast, weak sensitivity is observed near the four corners and in the immediate vicinity of the stagnation point, indicating that these regions are less influential in triggering the instability.

\subsection{Energy-budget analysis}  \label{subsec:energyBudget}
To identify the energetic origins of the symmetry-breaking instability
and further substantiate its physical mechanism, we perform a perturbation energy-budget analysis following \citet{sadanandan2004global,spyridakis_viscoelastic_2024}. 
By integrating over the domain $\Omega$, the inner product of the linearized momentum equation with the perturbation velocity $\bsu'$, we derive the evolution equation for the perturbation energy $E$ associated with polymeric stress as detailed in Appendix \ref{appen:energy}:
\begin{equation} \label{eq:energy_def}
    E = \frac{Wi}{2} \int_{\Omega} \bsu' \bcdot \bnabla \bcdot \bstau' \dd \Omega.
\end{equation}
Strictly speaking, $E$ is the perturbation energy per unit
out-of-plane length, but is referred to simply as the perturbation energy for brevity throughout this 2D study.
Notably, its rate of change ${\rm d}E/{\rm d}t$ quantifies whether 
perturbations are amplified or attenuated: positive rates indicate temporal growth of the perturbation energy and therefore instability.
To gain a deeper understanding, we further decompose this instability indicator into four energy-budget terms, 
\begin{equation} \label{eq:energy_equation}
	\frac{{\rm d} E}{{\rm d} t} = \phi_{p} + \phi_{vis} + \phi_{\epsilon} + \phi_{coup},
\end{equation}
with each representing the rate of energy production/dissipation rooted in a specific physical mechanism. 
In particular, $\phi_{p}$ is the rate of work done by the perturbation pressure, $\phi_{vis}$ represents the viscous dissipation rate that is consistently negative, while $\phi_{\epsilon}$ accounts for the energy contribution due to polymer-stress diffusion, with their expressions provided in Appendix \ref{appen:energy}. 
The remaining term, $\phi_{coup}$, standing for the rate of energy exchange arising from the coupling between the base state and the perturbation, 
is defined as
\begin{equation}
    \phi_{coup} = -Wi \int_{\Omega} \bsu'\bcdot \bnabla \bcdot \bs{\psi}' \dd \Omega,
\end{equation}
where 
\begin{equation}
    \bs{\psi}' = \bsu^{b} \bcdot \bnabla \bstau' + \bsu' \bcdot \bnabla \bstau^{b}-\bstau^{b} \bcdot \sfbiG'-\bstau' \bcdot \sfbiG^{b} - \left(\sfbiG^{b}\right)^{\trans} \bcdot \bstau' - \sfbiG^{\prime } \bcdot \bstau^{b}.
\end{equation}

\begin{figure}[!htbp]
    \centering
    \includegraphics[width=1.0\linewidth]{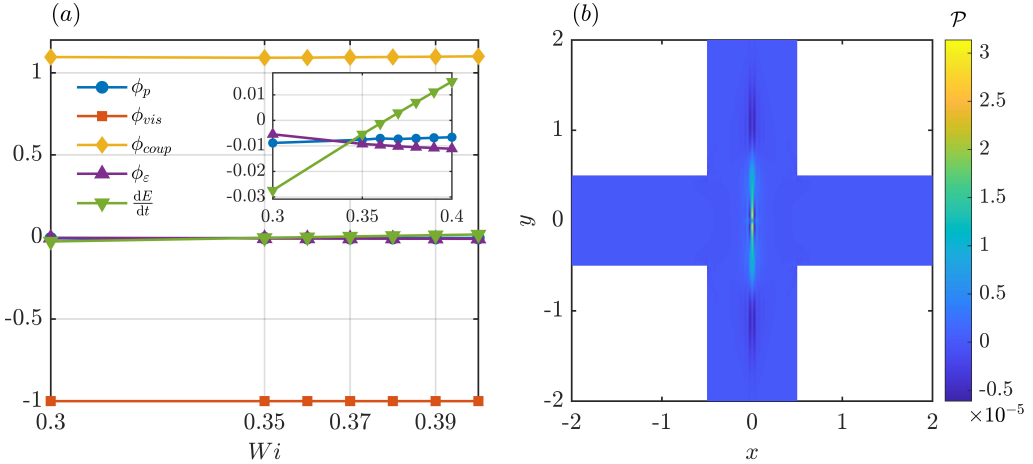}
    \caption{(a) Rate of change of the perturbation energy, $\dd E/\dd t$, and its four energy-budget terms versus $Wi$, see equation~\eqref{eq:energy_def}. (b) Distribution of the polymeric power density, $\mP$. Positive values indicate the energetic source that feeds the instability.}
    \label{fig:MN40_energyBudget}
\end{figure}

Figure~\ref{fig:MN40_energyBudget}(a) shows the 
perturbation-energy growth rate $\dd E/\dd t$ and its four energy-budget terms versus $Wi$, where  we have normalized all terms by $|\phi_{vis}|$ such that $\phi_{vis} = -1$ by construction. 
The net growth rate ${\rm d}E/{\rm d}t$ becomes positive from negative when $Wi>0.36$, 
consistent with the critical Weissenberg number of the bifurcation identified earlier (see figure~\ref{fig:DQ_eps5e-5_subfig}). Among all the energy-budget terms, $\phi_{coup}$ is the dominant positive one, and therefore, represents the major energy source.

To investigate the spatial distribution of the energy production rate, we first rewrite the integrand on the right-hand side of equation \eqref{eq:energy_def} as
\begin{equation}
    \bsu' \bcdot \bnabla \bcdot \bstau' = \grad \bcdot \left(\bstau' \bcdot \bsu'\right) - \frac{1}{2}\bstau'\bs{:} \left[ \grad \bsu' + (\grad \bsu')^{\trans} \right].
\end{equation}
Using the divergence theorem, we obtain
\begin{equation}\label{eq:div}
    \int_{\Omega} \grad \bcdot \left(\bstau' \bcdot \bsu'\right) \dd\Omega = \int_{\partial\Omega} \bs{n} \bcdot  \bstau' \bcdot \bsu' \dd \Gamma = 0,
\end{equation}
since $\htau\bcdot\hu = \bs{0}$ at all boundaries (as discussed in Appendix \ref{appen:zero_boundary}). 
Hence, this integral makes no net contribution to the perturbation energy growth. Then, we focus on the linear growth stage, when the perturbations $\bsu'$ and $\bstau'$ can be represented by the leading normal mode,
\begin{equation}
    \bsu' \sim e^{\sigma_r t} \hat{\bsu}, \quad \bstau'\sim e^{\sigma_r t} \hat{\bstau},
\end{equation}
where $\hat{\bsu}$ and $\hat{\bstau}$ represent the leading direct eigenmodes of the velocity and polymeric stress, respectively. Substituting into the perturbation energy and calculating its derivative leads to 
\begin{equation}
    \frac{\dd E}{\dd t} \sim \sigma_r e^{2\sigma_r t} Wi \, \int_{\Omega} \mP \dd \Omega,
\end{equation}
where 
\begin{equation} \label{eq:mp}
    \mP = - \htau \, \bs{:} \, \frac{ \bnabla\hu + \left(\bnabla\hu\right)^{\trans} }{2}
\end{equation}
denotes the perturbation polymeric power density (per unit volume), which characterizes the local rate of energy exchange between disturbance kinetic energy density (ED) and polymer elastic ED.
With this sign convention, $\mP$ represents the rate at which the disturbance polymeric stress performs work on the disturbance flow. Thus, negative $\mP$ corresponds to transfer from the kinetic ED of the disturbance fluid motion to the elastic ED stored in deformed polymer chains; whereas positive $\mP$ indicates the release of stored elastic ED back into the disturbance kinetic ED, which destabilizes the flow.

In the Newtonian limit, $Wi \to 0$, the perturbation polymeric stress becomes proportional to, and aligned with, the perturbation strain-rate tensor,
$\htau \to (1-\beta) \left[\bnabla\hu + (\bnabla\hu)^{\trans}\right]$. Consequently, $\mP$ is proportional to the negative of the disturbance viscous-dissipation density and is therefore non-positive throughout the domain, becoming strictly negative wherever the perturbation strain rate is nonzero. At larger $Wi$, the finite polymer relaxation time can introduce a local misalignment, or phase shift, between $\htau$ and $\bnabla\hu + (\bnabla\hu)^{\trans}$. Their inner product may then become negative in localized regions, corresponding to $\mP>0$, where stored elastic energy is released back to the disturbance flow and thus promotes instability.

As depicted in figure~\ref{fig:MN40_energyBudget}(b), the polymeric power density $\mP$ attains its largest positive values along a vertical ridge
aligning with the central zone of the extensional flows downstream of the stagnation point. 
This region marks the energetic source of the instability. Importantly, this ridge closely matches the stress-based wavemaker shown in figure~\ref{fig:MN40_wavemaker_Wi0.4}(b), hence confirming the spatial origin of the elastic instability.

\begin{figure}[!htbp]
    \centering
    \includegraphics[width=1\linewidth]{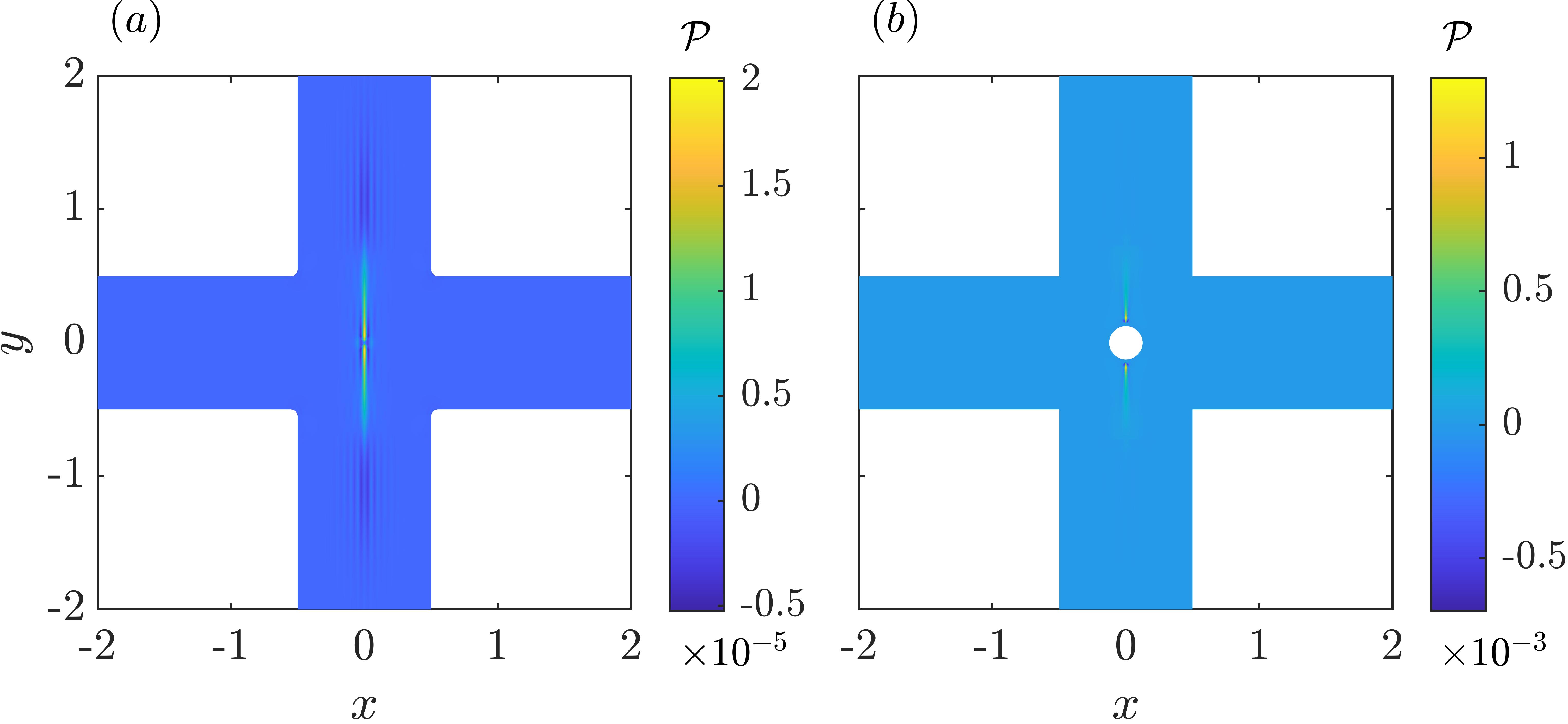}
    \caption{Distribution of the polymeric power density,
    $\mP$, for two geometric variants at $Wi=0.4$ and $\epsilon=0$. (a) Cross-slot with rounded corners of radius $0.05$. (b) Cross-slot with a circular cylinder of diameter $0.25$ at the center.}
    \label{fig:rc_cyl_tauDdotGradu_eps0}
\end{figure}

We conclude the energy-budget analysis by examining two geometric variants of the baseline cross-slot setting---rounded corners and a centered cylinder---to further elucidate the instability mechanism. 
Specifically, the first variant adopts rounded corners with  an arc radius of $0.05$, 
removing the geometric singularities of sharp corners and resulting stress singularities. 
The second variant introduces a circular cylinder of diameter $0.25$ at the center, thereby eliminating the original free stagnation point. 
For these two cases, we set $\epsilon=0$ owing to substantial numerical difficulties encountered at $\epsilon=5\times10^{-5}$. This choice does not provide a strictly equivalent benchmark comparison, but it is not expected to alter the qualitative findings or the resulting physical interpretation.
We perform LSA and energy-budget analysis, with the direct eigenmodes presented in Appendix \ref{appen:supple}.  
Figure~\ref{fig:rc_cyl_tauDdotGradu_eps0} shows the distribution of the polymeric power density $\mP$.
Similar to the baseline setting, the instability persists, and the density $\mP$ remains strongly localized along vertical ridges lying within the central zone of the extensional flows downstream of the stagnation point or of the cylinder. 
This similarity indicates that strong extensional flows and the ensuing polymer stretching
rather than the corner singularity or the central stagnation point, 
constitute the essential ingredient of the elastic instability in cross-slot devices. Hence, 
this understanding justifies the persistence of symmetry-breaking instability of viscoelastic flows through the OSCER variant of cross-slot channel that features smooth corners and a nearly homogeneous extensional flow~\citep{haward2013instabilities, haward2016elastic,cruz2018characterization}.
It also explains why the onset of instability remains barely changed when the stagnation point is eliminated by inserting a small cylinder there~\citep{davoodi2019control}: the extensional flows downstream of the cylinder, together with the associated  birefringent strand and phase-delay-induced positive polymeric power density, persist despite the absence of the free stagnation point.

\section{Conclusion and Discussion}\label{sec:conclusion}
In this work, we revisit the stability of two-dimensional creeping viscoelastic flow in a cross-slot channel, to clarify the mechanism underlying its first, steady symmetry-breaking instability. Although this instability was first observed experimentally nearly half a century ago~\citep{gardner1982photon} and reproduced numerically nearly two decades ago~\citep{poole2007purely}, its physical origin remains debated. To address this long-standing issue, we study inertialess cross-slot flow of a diffusive Oldroyd-B fluid by combining DNS, global LSA, structural sensitivity analysis, and energy-budget analysis, all implemented within the FEM library FreeFem++.

Our DNS captures the canonical symmetric-to-asymmetric flow transition through a pitchfork bifurcation at $Wi_{cr} \approx 0.36$. The LSA gives a consistent prediction of the critical Weissenberg number and of the temporal growth rates obtained from DNS. Moreover, the base state used in LSA and the leading direct eigenmode are recovered by the dominant zero-frequency and second DMD modes, respectively. These results collectively provide a stringent validation of our LSA implementation.

The leading velocity eigenmode from the LSA forms a centered, diamond-shaped vortical structure with closed streamlines, while the corresponding normal-polymeric-stress eigenmode exhibits antisymmetric parity. Together, they constitute a chiral linear perturbation that drives spontaneous symmetry breaking in the viscoelastic cross-slot flow by tilting and rotating the birefringent strand generated by the vertically oriented extensional flows. This mechanistic picture is further highlighted by visualizing the perturbation conformation tensor via spectral decomposition.

Beyond DNS and LSA, structural sensitivity analysis reveals a stress-based wavemaker region that marks the spatial core of the instability. This region, where the flow is highly sensitive to localized polymeric-stress perturbations, lies with the extensional-flow core and comprises vertical filaments arranged symmetrically about the stagnation point. This spatial configuration coincides with regions of strong perturbation conformation, naturally linking the wavemaker to the polymer-stretching/compression response.

We examine the growth of perturbation energy through a decomposition into four energy-budget contributions, each rooted in a distinct physical mechanism. The coupling between the base state and the perturbation fields is found to provide the dominant contribution. 
It can be further characterized by the disturbance polymeric power density $\mP$ measuring the rate at which the disturbance polymeric stress performs work on the disturbance flow. Notably, positive $\mP$ is most pronounced in a vertical ridge along the centerline extensional flows downstream of the stagnation point, identifying the energetic source of the elastic instability. Its spatial overlap with the stress-based wavemaker further establishes this narrow, extension-dominated zone as the instability core. 
We further test this insight by studying two geometric variants of the baseline setting: a cross-slot channel with rounded corners and another containing a centered cylinder eliminating the stagnation point.
In both cases, peak power density $\mP$ appears along vertical ridges aligned with the central 
high-extension-rate zones, reinforcing the proposed instability mechanism.

These findings help reconcile the debate over whether the instability originates from sharp corners or from the 
stagnation point. Studies on the OSCER variant of the cross-slot channel~\citep{haward2013instabilities, haward2016elastic,cruz2018characterization} suggest that sharp corners are not necessary for the instability, whereas the cylinder-inserted, stagnation-point-free setup investigated in \citet{davoodi2019control} points instead to corner-induced streamline curvature and high deformation rates. We find that neither the corner nor the free central stagnation point 
is the essential ingredient; rather, the spatial core of the instability lies in narrow high-extension-rate regions within the extensional flows. This explanation is more consistent with the instability mechanism of ``large normal stresses and gradients in the birefringence strand'' in extensional flow proposed by \citet{cruz2014new}. As a side note, we find no evidence that upstream compressional-flow disturbances are relevant to the instability onset, contrary to earlier hypotheses.

In summary, our results offer insights into the long-standing 
controversy over the physical origin of the creeping viscoelastic-flow instability in the cross-slot channel. More broadly, this study demonstrates the potential of integrating global stability, structural sensitivity, and energy-budget analyses to understand instability  in viscoelastic flows through complex geometries for which analytical base states are unavailable. Future work will extend the current framework to systematically investigate related viscoelastic flow instabilities involving other ingredients, including 3D perturbations and flows, other complex geometries, finite inertia, realistic rheological models, unsteady base states, capillary interfaces, and deformable structures.

\begin{bmhead}[Funding]
L.Z. acknowledges the support from the Singapore Ministry of Education Academic Research Fund Tier 2 (MOE-T2EP50122-0015) and Tier 1 grants (A-8003190-00-00). Computation of the work was performed on resources of the National Supercomputing Centre, Singapore (https://www.nscc.sg), as well as those provided by NUS IT via a grant (NUSREC-HPC-00001).
\end{bmhead}

\begin{bmhead}[Declaration of interests]
The authors report no conflict of interest.
\end{bmhead}

\begin{appen}
\clearpage

\section{Numerical methods for DNS} \label{appen:DNS}
Here, we describe the FEM formulation and the time-advancement scheme used for the DNS.
We first derive the weak form of the DEVSS-G formulation \eqref{eq:DEVSSG}. 
We use $\breve{(\bcdot)}$ to denote the test functions for the corresponding variables.
Multiplying the momentum equation \eqref{eq:DEVSSG_momentum} by the velocity test function $\tbu = \left( \breve{u}, \breve{v}\right)$, 
integrating over the domain $\Omega$, and applying integration by parts to the pressure and viscous-stress terms gives:
\begin{eqnarray}
    & &\int_{\Omega} \left\{
    - p \bnabla \bcdot \tbu  
    + \bnabla \tbu \bs{:} \left[\bnabla\bsu+ \left(\bnabla \bsu\right)^{\trans} \right] 
    -\tbu \bcdot \bnabla \bcdot \bstau  + \tbu \bcdot  \left[ \left(1-\beta\right)\bnabla \bcdot \left( \sfbiG+\sfbiG^\trans\right)\right]  \right\} \dd\Omega
    \nonumber\\
     & & +\int_{\partial \Omega} \tbu \bcdot \left\{ \bs{n} \bcdot \left[ p\mathsfbi{I}- [\bnabla \bsu+\left(\bnabla \bsu\right)^\trans] \right] \right\} \dd \Gamma
    = 0,
\end{eqnarray}
where $\Omega$ denotes the computational domain, $\partial\Omega$ denotes its boundary,
$\dd\Omega$ is the area element in $\Omega$, $\dd\Gamma$ is the line element on $\partial\Omega$,
and $\bs{n} = (n_x,n_y)$ denotes the unit normal on $\partial\Omega$, pointing from the fluid toward the exterior.

The boundary term vanishes on the walls and inlet, where Dirichlet conditions are imposed on the velocity and the corresponding velocity test function is zero.
We consider a fully developed flow for the outlet boundary. For the upper outlet in the cross-slot, $u=0, p=0$ and $\frac{\partial v}{\partial y}=0$. Then the integral on the outlet boundary becomes:
\begin{equation}
    \int_{\partial \Omega} \tbu \bcdot \left\{ \bs{n} \bcdot \left[p\mathsfbi{I}- [\bnabla \bsu+\left(\bnabla \bsu\right)^\trans]\right]\right\} \dd \Gamma
    = -\int_{\partial \Omega} \tbu \bcdot \left\{ \bs{n} \bcdot \left[\bnabla \bsu+\left(\bnabla \bsu\right)^\trans\right]\right\} \dd \Gamma.
\end{equation}
For the upper outlet boundary featuring $n_x=0, n_y=1$, 
we apply the boundary condition $u=0$ and hence its associated test function  
$\breve{u}=0$. 
Then the boundary term there vanishes:
\begin{equation}
      -\tbu \bcdot \left\{ \bs{n} \bcdot \left[\bnabla \bsu+\left(\bnabla \bsu\right)^\trans \right]\right\}
    = -\breve{u}_{\kappa} n_{\zeta} \left(\frac{\partial u_{\kappa}}{\partial x_{\zeta}} 
      + \frac{\partial u_{\zeta}}{\partial x_{\kappa}}\right)
    = -2 \breve{v} \frac{\partial v}{\partial y} = 0.
\end{equation}
So as it does at the lower outlet.

The weak form of the velocity gradient approximation $\sfbiG = \bnabla \bsu$ is
\begin{equation}
    \int_{\Omega} \tbG \bs{:} \left(\sfbiG - \bnabla \bsu\right) \dd\Omega =0.
\end{equation}
The weak form of the continuity equation is
\begin{equation}
   \int_{\Omega} \tp \bnabla \bcdot \bsu \dd\Omega= 0.
\end{equation}
The weak form of the constitutive equation is
\begin{eqnarray}\label{eq:weak_consti}
     &&
     \int_{\Omega} \ttau \bs{:} \left[\bstau
     +Wi\left(\frac{\partial \bstau}{\partial t} 
     + \bsu \bcdot \bnabla \bstau
     -\bstau \bcdot \sfbiG
     -\sfbiG^{\trans} \bcdot \bstau \right) 
     + \left(\beta-1\right) \left(\sfbiG+\sfbiG^{\trans} \right)\right]
     +\epsilon \bnabla \ttau \bs{:} \bnabla  \bstau
      \dd\Omega \nonumber \\
     &&\quad - \epsilon \int_{\partial\Omega}  \bs{n} \bcdot \left(\bnabla  \bstau \bs{:} \ttau\right) \, \dd \Gamma =0.
\end{eqnarray}

We first note that the boundary integral involving stress diffusion in equation~\eqref{eq:weak_consti}
vanishes on each part of $\partial\Omega$. 
At the walls, we have set $\epsilon = 0$ following \citet{beneitez2025linear}, 
leading to zero integral.
At the inlets, 
we impose the analytical polymeric stress profile as Dirichlet boundary conditions, so the test function $\ttau = \bs{0}$, again rendering the integral zero.
At the outlets, the natural boundary condition of the weak formulation implicitly enforces $\bs{n} \bcdot \bnabla\bstau = \bs{0}$, so that
\begin{equation*}
    -\epsilon \int_{\partial\Omega} \bs{n} \bcdot \left(\bnabla\bstau \bs{:} \ttau\right)\,\mathrm{d}A
    = -\epsilon \int_{\partial\Omega} \left(\bs{n} \bcdot \bnabla\bstau\right) \bs{:} \ttau\,\mathrm{d}A = 0.
\end{equation*}
As the boundary contribution vanishes entirely, we obtain the following weak form of the constitutive equation:
\begin{equation}
     \int_{\Omega} \left\{
     \ttau \bs{:} \left[ \bstau
     +Wi\left(\frac{\partial \bstau}{\partial t} 
     + \bsu \bcdot \bnabla \bstau
     - \bstau \bcdot \sfbiG
     - \sfbiG^{\trans} \bcdot \bstau \right) 
     + \left(\beta-1\right)\left(\sfbiG+\sfbiG^{\trans} \right) \right]
     +\epsilon \bnabla \ttau : \bnabla  \bstau
      \right\} \dd\Omega = 0.
\end{equation}

For unsteady DNS, we advance the solution from time level $t_n$ at step $n$ to $t_{n+1}$ at step $n+1$, where $\Delta t = t_{n+1}-t_n$ denotes the step size. We employ a decoupled time-stepping scheme consisting of two substeps. 
In the first substep, given the stress field $\bstau_n$ at time level $t_n$, we solve a coupled system comprising the continuity equation, the momentum equation, and the auxiliary equation for the velocity-gradient tensor to obtain $p_{n+1}$, $\bsu_{n+1}$, and $\sfbiG_{n+1}$:
\begin{eqnarray}
    & &\int_{\Omega} \Big\{
    - p_{n+1} \bnabla \bcdot \tbu  + \bnabla \tbu \bs{:} \left[\bnabla \bsu_{n+1}+ \left( \bnabla \bsu_{n+1}\right)^\trans
    \right] 
    -\tbu \bcdot \bnabla \bcdot \bstau_n + \tbu \bcdot  \left[ \left(1-\beta\right)\bnabla \bcdot \left( \sfbiG_{n+1}+\sfbiG_{n+1}^\trans\right)\right] \nonumber\\
     & & \quad  -\tp \bnabla \bcdot \bsu_{n+1} +\tbG \bs{:} \left(\sfbiG_{n+1} -\bnabla \bsu_{n+1}\right) \Big\} \dd\Omega = 0.
\end{eqnarray}
In the second substep, the constitutive equation is solved to update $\bstau_{n+1}$ by a characteristic-Galerkin method~\citep{pironneau1982transport, machmoum2001finite,lukavcova2016energy}, namely,
\begin{eqnarray}
    &&\int_{\Omega} \Bigl\{
    \ttau \bs{:} \left[\bstau_{n+1}
    +Wi\left(
    \frac{\bstau_{n+1}-\bstau_n \circ \bs{X}_n(t_n)}{\Delta t}
    - \bstau_{n+1} \bcdot \sfbiG_{n+1}
    - \sfbiG_{n+1}^{\trans} \bcdot \bstau_{n+1} \right) \right. \nonumber \\
     && \left. \quad +\left(\beta-1\right)\left(\sfbiG_{n+1}+\sfbiG_{n+1}^{\trans}\right) \right]  
     +\epsilon \bnabla \ttau \bs{:} \bnabla \bstau_{n+1} \Bigr\} \dd\Omega
    = 0,
\end{eqnarray}
where $\bs{X}_n$ denotes the trajectory (or characteristics) of a fluid particle. 
$\bs{X}_n(t,\bs{x})$ is obtained by solving
\begin{equation}
    \begin{cases}
        \frac{\dd}{\dd t} \bs{X}_n \left(t,\bs{x}\right) = \bsu_{n}\bigl(\bs{X}_n(t,\bs{x})\bigr), \quad t\in[t_n,t_{n+1}],
        \\
         \bs{X}_n\left(t_{n+1},\bs{x}\right) = \bs{x}.
    \end{cases}
\end{equation}
Approximating this trajectory by a first-order scheme gives the upwind point $\bs{X}_n(t_n,\bs{x}) \approx \bs{x}-\bsu_n(\bs{x})\,\Delta t$. 
Thus the composition term is written as $\bstau_n\circ \bs{X}_n(t_n,\bs{x})=\bstau_n\bigl(\bs{X}_n(t_n, \bs{x})\bigr)$, which is the previous-step stress evaluated at the upwind point.

Finally, we spatially discretize the weak forms using the 
finite-element spaces introduced in 
\S\,\ref{sec:numerical}: triangular Taylor--Hood elements ($P_2$--$P_1$) for the velocity--pressure pair and continuous piecewise-linear elements ($P_1$) for $\bstau$ and $\sfbiG$. We then assemble and solve the resulting discrete systems in FreeFem++.
Within this discrete setting, we further stabilize the stress update with an SUPG scheme~\citep{baaijens1998mixed}, replacing the stress test function $\ttau$ by
\begin{equation}
    \ttau_{SUPG}=\ttau+\alpha \, \bsu \bcdot \bnabla \ttau,
\end{equation}
where $\alpha=c_s h/|\bsu|$, $h$ is the characteristic element size, and $c_s=0.3$ is a tunable constant.

\section{Global LSA: derivation and implementation} \label{appen:LSA}
Here, we detail the derivation for the global LSA and describe the associated numerical implementation.
We begin with the base state, whose governing equations are obtained by dropping the temporal derivative of equation~\eqref{eq:DEVSSG}:
\begin{subequations}
\begin{align}
&\bnabla \bcdot \bsu^{b}=0, \\
& \bnabla p^{b} 
-\bnabla \bcdot \left[ \bnabla \bsu^{b} + \left(\bnabla \bsu^{b}\right)^{\trans} \right]
-\bnabla \bcdot \bstau^{b} 
+ \left(1-\beta\right) \bnabla \bcdot\left[\sfbiG^{b} + \left(\sfbiG^{b}\right)^{\trans}\right]
=\bs{0}, \\
&\bstau^{b}
+Wi\left[
    \bsu^{b} \bcdot \bnabla \bstau^{b}
    - \bstau^{b} \bcdot \sfbiG^{b}
    - \left( \sfbiG^{b} \right)^{\trans} \bcdot \bstau^{b} 
\right]
= \left(1-\beta\right)\left[\sfbiG^{b} + \left(\sfbiG^{b}\right)^{\trans}\right]
+ \epsilon \nabla^2 \bstau^{b}, 
\\
& \sfbiG^{b}=\bnabla \bsu^{b}.
\end{align}
\label{eq:OB-ss}
\end{subequations}
Subtracting equation~\eqref{eq:OB-ss} from equation~\eqref{eq:DEVSSG}, we obtain the perturbation equation:
\begin{subequations}
\begin{align}
& \bnabla \bcdot \bsu^{\prime}=0, \\
& \bnabla p^{\prime}
-\bnabla \bcdot \left[ \bnabla \bsu^{\prime} + \left(\bnabla \bsu^{\prime}\right)^{\trans} \right]
-\bnabla \bcdot \bstau^{\prime}
+\left(1-\beta\right) \bnabla \bcdot\left[\sfbiG^{\prime} + \left(\sfbiG^{\prime}\right)^{\trans}\right]
=\bs{0}, 
\label{eq:ptb_momentum}
\\
& \bstau^{\prime}
+Wi\left[
\frac{\partial \bstau^{\prime}}{\partial t}
+\bsu^{b} \bcdot \bnabla \bstau^{\prime}
+\bsu^{\prime} \bcdot \bnabla \bstau^{b}
-\bstau^{b} \bcdot \sfbiG^{\prime}
-\bstau^{\prime} \bcdot \sfbiG^{b}
-\left(\sfbiG^{b}\right)^{\trans} \bcdot \bstau^{\prime}
-\left(\sfbiG^{\prime}\right)^{\trans} \bcdot \bstau^{b} 
\right] \nonumber \\
& \quad
=\left(1-\beta\right)\left[\sfbiG^{\prime} + \left(\sfbiG^{\prime}\right)^{\trans}\right] 
+ \epsilon \nabla^2 \bstau^{\prime},
\label{eq:ptb_constitutive}
\\
& \sfbiG^{\prime}=\bnabla \bsu^{\prime}.
\end{align}\label{eq:ptb_all}
\end{subequations}
Substituting the normal-mode form~\eqref{eq:normal_mode_ansatz} into equation~\eqref{eq:ptb_all}, 
we obtain the direct global stability eigenvalue problem~\eqref{eq:lsa}.
The corresponding coupled weak formulation is
\begin{eqnarray}
\label{eq:lsa_weak}
    & & \int_{\Omega} \sigma \ttau \bs{:} \hat{\bstau} \dd \Omega 
    = \int_{\Omega} \Bigl\{
    - \hat{p} \bnabla \bcdot \tbu 
    + \bnabla \tbu \bs{:} \left[\bnabla \hat{\bsu} + \left(\bnabla \hat{\bsu}\right)^{\trans} \right] 
    -\tbu \bcdot \bnabla \bcdot \hat{\bstau}  
    + \left(1-\beta\right) \tbu \bcdot   \bnabla \bcdot\left(\hat{\sfbiG}+\hat{\sfbiG}^{\trans}\right) 
    - \tp \bnabla \bcdot \hat{\bsu}
     \nonumber\\
     & & +\tbG \bs{:} \left(\hat{\sfbiG} -\bnabla \hat{\bsu}\right) 
     + \frac{1}{Wi}  \ttau \bs{:} \left[(1-\beta)\left(\hat{\sfbiG}+\hat{\sfbiG}^{\trans}\right) 
     -\hat{\bstau} \right] 
     - \frac{\epsilon}{Wi}  \bnabla \ttau \bs{:} \bnabla  \hat{\bstau}
     \nonumber \\
     & & -\ttau \bs{:} \left[
     \bsu^{b} \bcdot \bnabla \hat{\bstau}
     +\hat{\bsu} \bcdot \bnabla \bstau^{b}
     -\bstau^{b} \bcdot \hat{\sfbiG}
     -\hat{\bstau} \bcdot \sfbiG^{b}
     -\left(\sfbiG^{b}\right)^{\trans} \bcdot \hat{\bstau}
     -\hat{\sfbiG}^{\trans} \bcdot \bstau^{b} 
     \right] 
     \Bigl\} \dd\Omega.
\end{eqnarray}

Upon finite-element discretizations, 
we obtain the generalized matrix eigenvalue problem
\begin{equation} 
    \mathsfbi{A} \bs{z} = \sigma \mathsfbi{B} \bs{z} ,
    \label{Eq:GEignSys}
\end{equation}
where $\bs{z}$ is the discrete eigenvector,
$\mathsfbi{A}$ is the stiffness matrix assembled from the terms on the right hand side of equation~\eqref{eq:lsa_weak},
and $\mathsfbi{B}$ is the mass matrix arising from the term $\int_{\Omega} \ttau \bs{:} \hat{\bstau} \dd\Omega$ in the weak form. 
To solve the generalized eigenvalue problem \eqref{Eq:GEignSys}, we apply the shift-invert Arnoldi method that computes a limited set of eigenvalues near a specified shift value $\sigma_{shift}$, using the recurrence relation 
$\bs{z}_k = \left(\mathsfbi{A} - \sigma_{shift} \mathsfbi{B}\right)^{-1} \mathsfbi{B}\, \bs{z}_{k-1}$, where $k= 1, 2, \ldots$ denotes the iteration count. 
We set $\sigma_{shift} = 0$ 
to target eigenvalues near the origin of the complex plane.
We solve the eigenvalue problem in the framework of FreeFem++ \citep{hecht2012new} through its interface to the Scalable Library for Eigenvalue Problem Computations (SLEPc) \citep{hernandez2005slepc}.
The resulting eigenvector $\bs{z}$ is normalized by its $\ell^2$-norm, resulting in the direct eigenmodes presented in the main text.

We solve the steady-state nonlinear system $F\left(\bsu^{b}, p^{b}, \sfbiG^{b}, \bstau^{b}\right) = 0$ for the base state using Newton--Raphson method. 
The nonlinear residual $F$ is defined as
\begin{eqnarray}
    & &F\left(\bsu,p,\sfbiG,\bstau\right) =  \nonumber \\
    & &\int_{\Omega} \Bigl\{
    - p \bnabla \bcdot \tbu  
    + \bnabla \tbu \bs{:} \left[\bnabla \bsu + \left(\bnabla \bsu\right)^{\trans} \right] 
    -\tbu \bcdot \bnabla \bcdot \bstau 
    + \left(1-\beta\right) \tbu \bcdot  \left( \bnabla \bcdot \sfbiG + \bnabla \bcdot \sfbiG^\trans\right) 
    \nonumber \\
    & &-\tp \bnabla \bcdot \bsu 
    +\tbG \bs{:} \left(\sfbiG -\bnabla \bsu\right) 
     \nonumber\\
     & & + \ttau \bs{:} \left[\bstau
     +Wi\left(\bsu \bcdot \bnabla \bstau
     -\bstau \bcdot \sfbiG
     -\sfbiG^{\trans} \bcdot \bstau \right) 
     +\left(\beta-1\right)\left(\sfbiG+\sfbiG^{\trans} \right)\right]
     +\epsilon \bnabla \ttau \bs{:} \bnabla  \bstau
     \Bigl\} \dd\Omega, 
    \label{res_OB}
\end{eqnarray}
and the Jacobian $\delta F$ is
\begin{eqnarray}
    & & \delta F\left(\delta \bsu, \delta p, \delta \sfbiG, \delta \bstau\right) =  \nonumber \\
    & &\int_{\Omega} \Bigl\{
    - \delta p \bnabla \bcdot \tbu 
    + \bnabla \tbu \bs{:} \left[\bnabla \delta \bsu + \left(\bnabla \delta \bsu\right)^\trans \right] 
    -\tbu \bcdot \bnabla \bcdot \delta \bstau 
    + \left(1-\beta\right) \tbu \bcdot  
    \left[ 
        \bnabla \bcdot \delta \sfbiG
        +  \bnabla \bcdot \left(\delta \sfbiG\right)^\trans
    \right] \nonumber \\
    & &  -\tp \bnabla \bcdot \delta \bsu 
    + \tbG \bs{:} \left(\delta \sfbiG -\bnabla \delta \bsu\right) 
    + \ttau \bs{:} \delta \bstau
    \nonumber \\
    & &+ Wi \, \ttau \bs{:} \left[
    \delta \bsu \bcdot \bnabla \bstau 
    + \bsu \bcdot \bnabla \delta \bstau
    -\delta\bstau \bcdot \sfbiG  
    -\bstau \bcdot \delta\sfbiG 
    - \left(\delta \sfbiG\right)^{\trans} \bcdot \bstau  
    - \sfbiG^{\trans} \bcdot   \delta\bstau 
    \right] \nonumber \\
    & & + \left(\beta-1\right)\ttau \bs{:} \left[ \delta\sfbiG + \left(\delta\sfbiG\right)^{\trans} \right]
    +\epsilon \bnabla \ttau \bs{:} \bnabla  \delta \bstau
    \Bigl\} \dd\Omega. 
    \label{deltaF_OB}
\end{eqnarray} 
Discretizing on the same finite-element spaces as in Appendix~\ref{appen:DNS}, equations \eqref{res_OB}–\eqref{deltaF_OB} yield the algebraic residual system and its Jacobian, which we solve by Newton iterations in FreeFem++ \citep{hecht2012new} through the Scalable Nonlinear Equations Solvers (SNES) module of PETSc \citep{snes2026userguide}.
At each SNES iteration, the Jacobian $\delta F$ in equation~\eqref{deltaF_OB} is assembled and passed to a preconditioned Krylov linear solver, and a 
line-search procedure is applied to ensure convergence. The Newton iteration is terminated according to the default convergence criteria of PETSc's SNES module, namely, the relative residual tolerance is less than $10^{-8}$.
The initial guess can be seeded from DNS results.
To compute high-$Wi$ base states, we employ a continuation strategy in which the converged solution at a lower $Wi$ provides the initial guess at the next $Wi$, thereby facilitating convergence.

\subsection{Dynamic mode decomposition (DMD)}\label{appen:dmd}
To characterize the instability directly from the simulation data, we apply DMD~\citep{schmid2010dynamic} to the early-time transient of the DNS. 
DMD is a data-driven, equation-free technique that decomposes a sequence of flow snapshots into spatial modes, each evolving at a single complex rate \citep{rowley2009spectral}. 
The temporal growth rate extracted purely from the data can provide an independent benchmark \citep{grilli2013transition} for the 
LSA predictions.

Specifically, we collect $k+1$ equally spaced snapshots of the state variable $\bsq = [p,\bsu,\sfbiG,\bstau]$ during the initial exponential growth.
Each snapshot is flattened into the discretized state vector $\cq_i \in \mathbb{R}^{m}$, sampled at $t^s_i = t^s_1 + (i-1)\Delta t^s$ (the superscript $^{s}$ denotes sampling), where $m$ is the total number of degrees of freedom and $\Delta t^s$ is the sampling interval.
The snapshots are arranged into two time-shifted data matrices
\begin{equation}\label{eq:dmd_data}
    \cQ = \begin{bmatrix} \cq_1 & \cq_2 & \cdots & \cq_k \end{bmatrix}, \qquad
    \cQ_s = \begin{bmatrix} \cq_2 & \cq_3 & \cdots & \cq_{k+1} \end{bmatrix},
\end{equation}
so that the columns of $\cQ_s$ are those of $\cQ$ advanced by one sampling interval $\Delta t^s$.
DMD assumes that a linear operator $\bsK$ approximately maps each snapshot to the next, $\cq_{i+1} \approx \bsK\,\cq_i$, or
\begin{equation}\label{eq:dmd_operator}
    \cQ_s \approx \bsK\,\cQ,
    \qquad
    \bsK = \cQ_s\,\cQ^{+},
\end{equation}
where $\cQ^{+}$ is the Moore--Penrose pseudoinverse.

Because $m$ is large, $\bsK$ is not formed explicitly.
Instead, using the rank-$\rSVD$ truncated singular value decomposition $\cQ \approx \bsU\,\bs{\Sigma}\,\bsV^{\trans}$, with $\bsU \in \mathbb{R}^{m\times \rSVD}$, $\bs{\Sigma} \in \mathbb{R}^{\rSVD \times \rSVD}$ and $\bsV \in \mathbb{R}^{k\times \rSVD}$, 
$\bsK$ is represented in the rank-$\rSVD$ proper orthogonal decomposition subspace:
\begin{equation}\label{eq:dmd_reduced}
    \tilde{\bsK} = \bsU^{\trans}\,\bsK\,\bsU
    = \bsU^{\trans}\,\cQ_s\,\bsV\,\bs{\Sigma}^{-1}.
\end{equation}
We then solve the eigenvalue problem for $\tilde{\bsK}$,
\begin{equation}\label{eq:dmd_eig}
    \tilde{\bsK}\,\bsW = \bsW\,\bs{\Lambda},
\end{equation}
to obtain the eigenvector matrix $\bsW=[\bs{w}_1,\ldots,\bs{w}_\rSVD]$ and the diagonal matrix $\bs{\Lambda}=diag(\Lambda_1,\ldots,\Lambda_\rSVD)$ of discrete-time DMD eigenvalues.
Next, we construct the matrix of exact-DMD modes as
\begin{equation}\label{eq:dmd_modes}
    \bs{\Xi} = \cQ_s\,\bsV\,\bs{\Sigma}^{-1}\,\bsW,
\end{equation}
where $\bs{\Xi} = [\bs{\xi}_1,\ldots,\bs{\xi}_\rSVD]$ is the matrix whose columns are DMD eigenvectors $\bs{\xi}_j$.
Introducing the matrix of continuous-time DMD eigenvalues $\bs{\omega}^{DMD}=diag(\omega^{DMD-1},\ldots,\omega^{DMD-\rSVD})$ with
\begin{equation}\label{eq:dmd_omega}
    \bs{\omega}^{DMD} = \frac{\ln \bs{\Lambda}}{\Delta t^s},
\end{equation}
the state is reconstructed as
\begin{equation}\label{eq:dmd_recon}
    \cq(t) \approx \bs{\Xi}\,e^{\bs{\omega}^{DMD}(t-t^s_1)}\check{\bs{b}},
\end{equation}
where $\check{\bs{b}}=\bs{\Xi}^{+} \cq_1$ is the vector of DMD mode amplitudes obtained by a least-squares fit to the initial condition.

We use the exact-DMD implementation of the open-source PyDMD package \citep{ichinaga2024pydmd} to compute the DMD modes and their eigenvalues.
The real and imaginary parts of the continuous-time eigenvalue $\omega^{DMD-j}$ of the $j$th mode indicate the temporal growth rate and angular frequency, respectively.
The comparison with the LSA eigenmodes and eigenvalues is presented in \S\,\ref{subsec:lsa}.

\section{Adjoint problem} \label{appen:adjoint}
We derive the adjoint linearized equation for the sensitivity analysis presented in \S\,\ref{sec:sensitivity}. 
We start from the governing equations written compactly as $\mathcal{N}_{\bsQ} = 0$, where $\bsQ=(p,\bsu,\bstau)$ collects the state variables and $\mathcal{N}_{\bsQ}$ is defined as:
\begin{equation}
\mathcal{N}_{\bsQ} =
\begin{cases}
\bnabla \bcdot \bsu &\\
\beta \bnabla \bcdot  \left[ \bnabla \bsu + \left(\bnabla \bsu\right)^{\trans} \right] + \bnabla \bcdot \bstau - \bnabla p &\\
\bstau + Wi\left[
\frac{\partial \bstau}{\partial t} +
\left(\bsu \bcdot \bnabla\right)\bstau - \left(\bnabla \bsu\right)^{\trans} \bcdot \bstau - \bstau \bcdot \bnabla \bsu \right] - \left(1-\beta\right) \left[ \bnabla \bsu + \left(\bnabla \bsu\right)^{\trans} \right]
- \epsilon \nabla^2 \bstau.
&
\end{cases}
\end{equation}

Linearizing $\mathcal{N}_{\bsQ}$ about the base state $\bsQ^{b}=\left(p^{b},\bsu^{b},\bstau^{b}\right)$ and retaining only first-order terms in the perturbation $\bsQ^{\prime}=\left(p^{\prime},\bsu^{\prime},\bstau^{\prime}\right)$ yields:
\begin{equation}
\mathcal{N}_{\bsQ}^{\prime} = \begin{cases}
\bnabla \bcdot \bsu^{\prime} 
\\
 \beta \bnabla \bcdot \left[ \bnabla \bsu^{\prime}+\left(\bnabla \bsu^{\prime}\right)^{\trans} \right] + \bnabla \bcdot \bstau^{\prime} - \bnabla p^{\prime} \\
\bstau^{\prime} 
+ Wi \left[\left(\bsu^{\prime} \bcdot \bnabla\right)\bstau^{b} - \left(\bnabla \bsu^{\prime}\right)^{\trans} \bcdot \bstau^{b} - \bstau^{b} \bcdot \bnabla \bsu^{\prime} \right]  \ldots
\\
+ Wi \left[\left(\bsu^{b} \bcdot \bnabla\right)\bstau^{\prime} - \left(\bnabla \bsu^{b}\right)^{\trans} \bcdot \bstau^{\prime} - \bstau^{\prime} \bcdot \bnabla \bsu^{b} \right] 
- \left(1-\beta\right) \left[ \bnabla \bsu^{\prime}+\left(\bnabla \bsu^{\prime}\right)^{\trans} \right] +Wi \frac{\partial \bstau^{\prime}}{\partial t}
- \epsilon \nabla^2 \bstau^{\prime}.
\end{cases}
\end{equation}

The adjoint state $\bsQ^{\dagger}=\left(p^{\dagger},\bsu^{\dagger},\bstau^{\dagger}\right)$ is defined by requiring that the inner product $\langle \bsQ^{\dagger},\mathcal{N}_{\bsQ}^{\prime}\rangle=0$ for every admissible perturbation $\bsQ^{\prime}$, with the $L^2$ inner product defined as
\begin{equation}
    \langle  \bs{a} ,  \bs{b} \rangle 
    = \int_{\Omega} \bs{a} \bcdot  \bs{b} \ \dd \Omega, \quad
    \langle  \mathsfbi{P} ,  \mathsfbi{S} \rangle 
    = \int_{\Omega} \mathsfbi{P} \bs{:} \mathsfbi{S} \ \dd \Omega,
\end{equation}
where $\bs{a}$, $\bs{b}$ are vectors, and $\mathsfbi{P}$, $\mathsfbi{S}$ are second-order tensors.

Expanding $\langle \bsQ^{\dagger},\mathcal{N}_{\bsQ}^{\prime}\rangle$ explicitly produces the weak form:
\begin{eqnarray} 
0 = \langle \bsQ^{\dagger}, \mathcal{N}_{\bsQ}^{\prime} \rangle 
&=& 
  \beta \langle \bsu^{\dagger},  \bnabla \bcdot \left[ \bnabla \bsu^{\prime} + \left(\bnabla \bsu^{\prime}\right)^{\trans} \right] \rangle
+ \langle \bsu^{\dagger}, \bnabla \bcdot \bstau^{\prime} \rangle 
- \langle \bsu^{\dagger}, \bnabla p^{\prime} \rangle
+  \langle p^{\dagger}, \bnabla \bcdot \bsu^{\prime} \rangle \nonumber \\ 
&+& \langle \bstau^{\dagger}, \bstau^{\prime} \rangle 
+ Wi\langle \bstau^{\dagger}, \bsu^{\prime} \bcdot \bnabla\bstau^{b} \rangle 
-  Wi \langle \bstau^{\dagger}, \left(\bnabla\bsu^{\prime}\right)^{\trans} \bcdot \bstau^{b} \rangle
-  Wi \langle\bstau^{\dagger}, \bstau^{b} \bcdot \bnabla\bsu^{\prime} \rangle \nonumber \\ 
&+& Wi\langle \bstau^{\dagger}, \bsu \bcdot \bnabla\bstau^{\prime} \rangle  
- Wi\langle \bstau^{\dagger}, \left(\bnabla \bsu^{b}\right)^{\trans} \bcdot \bstau^{\prime} \rangle 
- Wi\langle \bstau^{\dagger}, \bstau^{\prime} \bcdot \bnabla \bsu^{b} \rangle \nonumber \\
&-& (1-\beta) \langle \bstau^{\dagger}, \bnabla \bsu^{\prime}+\left(\bnabla \bsu^{\prime}\right)^{\trans} \rangle 
+ Wi \langle \bstau^{\dagger}, \frac{\partial \bstau^{\prime}}{\partial t}\rangle
- \epsilon \langle \bstau^{\dagger}, \nabla^2 \bstau^{\prime} \rangle.
\end{eqnarray}
Integration by parts is then applied to each term so as to transfer all derivatives from $\bsQ^{\prime}$ onto $\bsQ^{\dagger}$. 
Since $\bsQ^{\prime}$ is arbitrary, the resulting domain integrand must vanish identically, which produces the adjoint linearized equation:
\begin{equation}
     \langle \bsQ^{\dagger}, \mathcal{N}^{\prime}_{\bsQ} \rangle
     = \langle \mathcal{N}^{\dagger}_{\bsQ^{\dagger}}, \bsQ^{\prime} \rangle + B.I.,
\end{equation}
where $B.I.$ stands for the boundary integral term, while the domain-integral contribution is
\begin{eqnarray} 
\label{eq:int_adjoint}
&&\langle \mathcal{N}^{\dagger}_{\bsQ^{\dagger}}, \bsQ^{\prime} \rangle 
=  \beta \langle \bnabla \bcdot  \left[\bnabla \bsu^{\dagger} + \left(\bnabla \bsu^{\dagger}\right)^{\trans}\right], \bsu^{\prime} \rangle
    - \langle \bnabla p^{\dagger}, \bsu^{\prime}\rangle
    + Wi \langle \bsu^{\prime} ,  \bnabla \bstau^{b} \bs{:} \bstau^{\dagger} \rangle
    + 2 Wi \langle \bsu^{\prime} , \bnabla \bcdot \left(\bstau^{b} \bcdot \bstau^{\dagger}\right) \rangle
    \nonumber \\
&&  + 2 \left(1-\beta\right) \langle \bsu^{\prime}, \bnabla \bcdot \bstau^{\dagger} \rangle
    - \langle p^{\prime} , \bnabla \bcdot \bsu^{\dagger} \rangle \nonumber \\
&&  + \langle \bstau^{\dagger}, \bstau^{\prime} \rangle
    - Wi \langle \bstau^{\prime}, \bsu^{b} \bcdot \bnabla \bstau^{\dagger} \rangle
    - Wi \langle \bstau^{\prime}, \bnabla \bsu^{b} \bcdot \bstau^{\dagger} \rangle
    - Wi \langle \bstau^{\prime},  \bstau^{\dagger} \bcdot \left(\bnabla \bsu^{b}\right)^{\trans} \rangle
    \\ \nonumber
&&  - \langle \frac{1}{2}  \left[\bnabla \bsu^{\dagger} + \left(\bnabla \bsu^{\dagger}\right)^{\trans}\right], \bstau^{\prime} \rangle 
    + Wi \frac{\partial}{\partial t} \langle \bstau^{\dagger}, \bstau^{\prime} \rangle 
    - Wi \langle \frac{\partial \bstau^{\dagger}}{\partial t}, \bstau^{\prime} \rangle
    - \epsilon \langle \bstau^{\prime} \bs{:} \nabla^2 \bstau^{\dagger} \rangle.
    \nonumber
\end{eqnarray}
In addition, the boundary integral term reads
\begin{eqnarray} 
B.I. =
&& 
\beta \int_{\partial \Omega} \bsu^{\dagger} \bcdot \left[\bnabla \bsu^{\prime} + \left( \bnabla \bsu^{\prime} \right)^{\trans}\right] \bcdot \bs{n} \, \dd \Gamma 
\nonumber \\ 
&& - \int_{\partial \Omega} \bs{n} \bcdot \bsu^{\dagger} p^{\prime} \, \dd \Gamma 
\nonumber \\  
&& + \int_{\partial \Omega} 
\bsu^{\prime} \bcdot \left\{
- \beta  \bs{n} \bcdot \left[\bnabla \bsu^{\dagger} + \left(\bnabla \bsu^{\dagger}\right)^{\trans}
\right]  
+ \bs{n}  p^{\dagger} 
- 2Wi \, \bs{n} \bcdot \left( \bstau^{b} \bcdot \bstau^{\dagger}\right) 
- 2\left(1-\beta\right)  \bs{n} \bcdot \bstau^{\dagger} 
\right\} \, \dd \Gamma 
 \nonumber \\
&& + \int_{\partial \Omega} 
 \bstau^{\prime} \bs{:} \left[
\bs{n} \bsu^{\dagger} + Wi  \left(\bs{n} \bcdot \bsu^{b}\right) \bstau^{\dagger}
\right] \, \dd \Gamma \nonumber \\
&& - \epsilon \int_{\partial\Omega}  \left[ \bs{n} \bcdot \left(\bnabla \bstau^{\prime} \bs{:} \bstau^{\dagger}\right) 
                              -\bs{n} \bcdot \left(\bnabla \bstau^{\dagger} \bs{:} \bstau^{\prime}\right) \right] \dd \Gamma.
\end{eqnarray}

Requiring the domain integrand to vanish for arbitrary $\bsQ^{\prime}$ in equation~\eqref{eq:int_adjoint} gives the following adjoint linearized equation:
\begin{subequations}
\label{eq:adjoint_apd}
\begin{align}
& \bnabla \bcdot \bsu^{\dagger} = 0, \\
&\beta \bnabla \bcdot \left[ \bnabla \bsu^{\dagger} +  \left( \bnabla \bsu^{\dagger} \right)^{\trans} \right]
- \bnabla p^{\dagger} 
+ Wi \bnabla\bstau^{b} \bs{:} \bstau^{\dagger}
+ 2Wi\bnabla \bcdot \left(\bstau^{b} \bcdot \bstau^{\dagger}\right) 
 + 2\left(1-\beta\right) \bnabla \bcdot \bstau^{\dagger} = \bs{0}, \\
& \bstau^{\dagger} - Wi\left[  
  \frac{\partial \bstau^{\dagger}}{\partial t} 
+ \bsu^{b} \bcdot \bnabla\bstau^{\dagger} 
+ \bnabla \bsu^{b} \bcdot \bstau^{\dagger} 
+ \bstau^{\dagger} \bcdot \left(\bnabla \bsu^{b}\right)^{\trans}\right] 
- \frac{1}{2}\left[ \bnabla \bsu^{\dagger} +  \left( \bnabla \bsu^{\dagger} \right)^{\trans} \right] 
- \epsilon \nabla^2 \bstau^{\dagger}
= \bs{0}. 
\end{align} 
\end{subequations}
In the spirit of the DEVSS-G formulation, 
equation~\eqref{eq:adjoint_apd} can be reformulated to
\begin{subequations} \label{eq:adjointG_apd}
\begin{align}
    &\bnabla \bcdot \bsu^{\dagger} = 0, \\
    & \bnabla\bcdot\left[ \bnabla \bsu^{\dagger} + \left(\bnabla \bsu^{\dagger}\right)^{\trans}\right] 
    + (\beta-1)\bnabla \bcdot \left[ \sfbiG^{\dagger}+ \left(\sfbiG^{\dagger}\right)^{\trans} \right] - \bnabla p^{\dagger} + Wi  \bnabla\bstau^{b} : \bstau^{\dagger}  \nonumber \\
    & \quad + 2Wi\bnabla \bcdot \left(\bstau^{b} \bcdot \bstau^{\dagger}\right) + 2\left(1-\beta\right) \bnabla \bcdot \bstau^{\dagger} = \bs{0},  \\
    &\bstau^{\dagger} 
    - Wi\left[ \frac{\partial \bstau^{\dagger}}{\partial t} 
    + \bsu^{b} \bcdot \bnabla\bstau^{\dagger} 
    + \bnabla \bsu^{b} \bcdot \bstau^{\dagger} 
    + \bstau^{\dagger} \bcdot \left(\bnabla \bsu^{b}\right)^{\trans} \right] 
    - \frac{1}{2} \left[\sfbiG^{\dagger} 
    + \left(\sfbiG^{\dagger}\right)^{\trans}\right] 
    - \epsilon \nabla^2 \bstau^{\dagger} = \bs{0}, \\
    & \sfbiG^{\dagger} = \bnabla \bsu^{\dagger}.
\end{align} 
\end{subequations}
The boundary conditions of the adjoint linearized equation are obtained by requiring the boundary integral $B.I.$ to vanish for arbitrary admissible $\bsQ^{\prime}$.
At the walls, we impose 
\begin{equation} \label{eq:adj_bc1}
    \bsu^{\dagger}=\bs{0}.
\end{equation}
At the inlets, the boundary conditions are 
\begin{equation} \label{eq:adj_bc2}
    \bsu^{\dagger}=\bs{0}, \quad \bstau^{\dagger}=\bs{0}.
\end{equation}
At the outlets, the boundary conditions are
\begin{subequations} \label{eq:adj_bc3}
    \begin{align}
        & \bs{n} \bcdot \bnabla \bstau^{\dagger}=\bs{0}, \label{adj-outBC-a}
        \\
        & - \beta  \bs{n} \bcdot   \left[ \bnabla \bsu^{\dagger} +  \left( \bnabla \bsu^{\dagger} \right)^{\trans} \right]  
        + \bs{n}  p^{\dagger} 
        - 2Wi  \, \bs{n} \bcdot \left(\bstau^{b} \bcdot \bstau^{\dagger}\right)  
        - 2(1-\beta)  \bs{n} \bcdot \bstau^{\dagger} =\bs{0},
        \label{adj-outBC-b}
        \\
        & \bs{n} \bsu^{\dagger} + Wi \, \left(\bs{n} \bcdot \bsu^{b}\right) \bstau^{\dagger}= \bs{0}
        \label{adj-outBC-c}.
    \end{align}
\end{subequations}
Substituting the normal-mode form $\bsq^{\dagger}(\bs{x},t) = e^{-\sigma^{\dagger} t}\hat{\bsq}^{\dagger}(\bs{x})$ into equation~\eqref{eq:adjointG_apd}, we obtain the adjoint eigenvalue problem~\eqref{eq:adj_eigen}.
The boundary conditions for \eqref{eq:adj_eigen} can be obtained by substituting the normal-mode form into equations~\eqref{eq:adj_bc1}, \eqref{eq:adj_bc2}, and \eqref{eq:adj_bc3}. 
We formulate the weak form of the adjoint eigenvalue problem by multiplying equation~\eqref{eq:adj_eigen} by the test functions and integrating over $\Omega$.
Integration by parts generates boundary integrals, through which the boundary conditions of the adjoint eigenvalue problem are imposed, leading to the weak-form adjoint eigenvalue problem:
\begin{eqnarray} \label{eq:eigs_adj}
    & & \int_{\Omega} -\sigma^{\dagger} \ttau \bs{:} \hat{\bstau}^{\dagger} \dd\Omega
    =
    \int_{\Omega} \Biggl\{ 
      \hat{p}^{\dagger} \bnabla \bcdot \tbu 
    - \bnabla \tbu \bs{:} \left[\bnabla \hat{\bsu}^{\dagger} + \left(\bnabla \hat{\bsu}^{\dagger}\right)^{\trans} \right] 
    + (1-\beta) \bnabla \tbu \bs{:}  \left[\hat{\sfbiG}^{\dagger}+\left(\hat{\sfbiG}^{\dagger}\right)^{\trans}\right]
    + Wi \, \tbu \bcdot \left(\bnabla \bstau^{b} \bs{:} \hat{\bstau}^{\dagger}\right) \nonumber \\ 
    && - Wi  \, \hat{\bstau}^{\dagger} \bs{:} \left[\bstau^{b} \bcdot \bnabla\tbu 
                            + (\bnabla \tbu)^{\trans}  \bcdot \bstau^{b} \right]
      - \left(1-\beta\right) \hat{\bstau}^{\dagger} \bs{:}  \left[\bnabla \tbu + \left(\bnabla \tbu\right)^{\trans}\right]
     -\tp \bnabla \bcdot \hat{\bsu}^{\dagger} +\tbG \bs{:} \left(\hat{\sfbiG}^{\dagger} -\bnabla \hat{\bsu}^{\dagger}\right) 
     \nonumber \\
     & & + \ttau \bs{:} \left[
     \frac{\hat{\bstau}^{\dagger}}{Wi} 
     - \sfbiG^{b} \bcdot \hat{\bstau}^{\dagger}
     - \hat{\bstau}^{\dagger} \bcdot \left(\sfbiG ^{b}\right)^{\trans}
     \right]
     + \frac{1}{Wi} \hat{\bsu}^{\dagger} \bcdot \bnabla \bcdot \ttau
     + \hat{\bstau}^{\dagger} \bs{:}  \left[\bsu^{b} \bcdot \bnabla \ttau\right]  
     +  \frac{\epsilon}{Wi} \bnabla \ttau \bs{:} \bnabla \hat{\bstau}^{\dagger} \,
     \Biggl\} \dd\Omega
     \nonumber \\
     & & -\int_{\partial \Omega }
     \ttau\bs{:} \left(
         \bs{n} \bcdot \bsu^{b} \hat{\bstau}^{\dagger}
     \right) \, \dd \Gamma. 
\end{eqnarray}

Discretized on the same finite-element spaces as in Appendix~\ref{appen:LSA}, the adjoint eigenvalue problem~\eqref{eq:eigs_adj} yields the generalized matrix eigenvalue problem $\mathsfbi{A}^{\dagger}\bs{z}^{\dagger}=\sigma^{\dagger}\mathsfbi{B}^{\dagger}\bs{z}^{\dagger}$, 
which we solve with the same shift–invert Arnoldi method through SLEPc as in Appendix~\ref{appen:LSA}.
The adjoint eigenvector $\bs{z}^\dagger$ is normalized by its $\ell^2$-norm, leading to the adjoint eigenmodes shown in the main text.

\section{Energy-budget analysis} \label{appen:energy}
Here, we derive the perturbation energy budget discussed in 
\S\,\ref{subsec:energyBudget}.
We first rewrite the constitutive equation for perturbation \eqref{eq:ptb_constitutive} as
\begin{equation}
    \bstau^{\prime} 
    = \left(1-\beta\right)\left[\sfbiG^{\prime} + \left(\sfbiG^{\prime}\right)^{\trans}\right]
    + \epsilon \nabla^2 \bstau^{\prime}
    - Wi \left( \frac{\partial \bstau^{\prime}}{\partial t} + \bs{\psi}^{\prime} \right),
    \label{eq:btaup}
\end{equation}
where $\bs{\psi}^{\prime} $ denotes the coupling terms of the base state and perturbation,
\begin{equation}
   \bs{\psi}^{\prime} 
   = \bsu^{b} \bcdot \bnabla \bstau^{\prime}
   +\bsu^{\prime} \bcdot \bnabla \bstau^{b}
   -\bstau^{b} \bcdot \sfbiG^{\prime}
   -\bstau^{\prime} \bcdot \sfbiG^{b}
   -\left(\sfbiG^{b}\right)^{\trans} \bcdot \bstau^{\prime}
   -\left(\sfbiG^{\prime}\right)^{\trans} \bcdot \bstau^{b}.
   \label{Eq_bpsi}
\end{equation}
Substituting equation~\eqref{eq:btaup} 
into equation~\eqref{eq:ptb_momentum}, the linearized perturbation momentum equation becomes:
\begin{equation}
    Wi \frac{\partial}{\partial t} \left( \bnabla \bcdot \bstau^{\prime}\right) 
    = -\bnabla p^{\prime} 
    + \bnabla \bcdot\left[\sfbiG^{\prime} + \left(\sfbiG^{\prime}\right)^{\trans}\right] 
    + \epsilon \bnabla  \bcdot \left( \nabla^2 \bstau^{\prime}\right)
    -  Wi \bnabla \bcdot  \bs{\psi}^{\prime}.
    \label{eq:LNEQwPsi}
\end{equation}
To derive the evolution equation for the perturbation energy $E$ defined in equation~\eqref{eq:energy_def}, we take the scalar product of equation~\eqref{eq:LNEQwPsi} with $\bsu^{\prime}$ and integrate over the domain following \citet{sadanandan2004global}, which yields
\begin{equation}\label{eq:energy_evolve}
    \int_{\Omega} \frac{Wi}{2} \frac{\partial}{\partial t} \left( \bsu^{\prime} \bcdot \bnabla \bcdot \bstau^{\prime}\right) \dd\Omega
    = \int_{\Omega} \left[
    -\bsu^{\prime} \bcdot \bnabla p^{\prime} 
    + \bsu^{\prime} \bcdot \bnabla \bcdot\left[\sfbiG^{\prime}+\left(\sfbiG^{\prime}\right)^{\trans}\right]
    +  \epsilon  \bsu^{\prime} \bcdot  \bnabla \bcdot \left(\nabla^2 \bstau^{\prime}\right)
    -  Wi\, \bsu^{\prime} \bcdot \bnabla \bcdot  \bs{\psi}^{\prime}    
    \right] \dd\Omega.
\end{equation}
The left-hand side of equation~\eqref{eq:energy_evolve} represents the rate of the change of the perturbation energy $E$ (strictly speaking, perturbation energy per unit out-of-plan length),
\begin{align}\label{eq:dEdt}
    \frac{\dd E}{\dd t} = \int_{\Omega} \frac{Wi}{2}  \frac{\partial}{\partial t} \left( \bsu^{\prime} \bcdot \bnabla \bcdot \bstau^{\prime}\right) \dd\Omega = \frac{Wi}{2} \frac{\dd }{\dd t} \int_{\Omega}
         \bsu^{\prime} \bcdot \bnabla \bcdot \bstau^{\prime}
        \dd\Omega. 
\end{align}
Consistent with equation~\eqref{eq:energy_equation}, the right-hand side of equation~\eqref{eq:energy_evolve} can be decomposed into the following four energy-budget terms: 
\begin{subequations} \label{eq:energyApd}
\begin{align}
    & \phi_p = -\int_{\Omega}  \bsu^{\prime} \bcdot \bnabla p^{\prime}   \dd\Omega,
    \\
    &     \phi_{vis} = \int_{\Omega}  \bsu^{\prime} \bcdot \bnabla \bcdot\left[\sfbiG^{\prime} + \left(\sfbiG^{\prime}\right)^{\trans}\right]     \dd\Omega,
    \\
    &     \phi_{coup} = - Wi \int_{\Omega} \bsu^{\prime} \bcdot \bnabla \bcdot  \bs{\psi}^{\prime}
     \dd\Omega,
    \\
    &     \phi_{\epsilon} 
    = \epsilon \int_{\Omega}   \bsu^{\prime} \bcdot \bnabla \bcdot \left(\nabla^2 \bstau^{\prime}\right)   
     \dd\Omega.
\end{align}
\end{subequations}
Substituting the normal-mode form~\eqref{eq:normal_mode_ansatz} into equations~\eqref{eq:dEdt} and \eqref{eq:energyApd}, we obtain the corresponding modal forms for $\dd E/\dd t$ and the energy-budget terms:
\begin{subequations}
\begin{align}
    & \frac{\dd E}{\dd t} = \sigma  e^{2\sigma t}  Wi
        \int_{\Omega}
          \hu \bcdot \bnabla \bcdot \hat{\bstau}
        \dd\Omega,
    \\
    & \phi_p = - e^{2\sigma t} \int_{\Omega}  \hu \bcdot \bnabla \hat{p}   \dd\Omega,
    \\
    & \phi_{vis} = e^{2\sigma t} \int_{\Omega}  \hu \bcdot \bnabla \bcdot
      \left( \hat{\sfbiG} + \hat{\sfbiG}^{\trans} \right)     \dd\Omega,
    \\
    & \phi_{coup} = -  e^{2\sigma t}  Wi \int_{\Omega}   \hu \bcdot \bnabla \bcdot  \hat{\bs{\psi}}    
     \dd\Omega,
    \\
    & \phi_{\epsilon} 
    = \epsilon e^{2\sigma t}  \int_{\Omega}  \hat{\bsu} \bcdot \bnabla \bcdot \left(\nabla^2 \hat{\bstau}   
     \right) \dd\Omega.
\end{align}
\end{subequations}

\subsection{Regarding $\htau\bcdot\hu = \bs{0}$ at all boundaries}\label{appen:zero_boundary}
In \S\,\ref{subsec:energyBudget}, equation~\eqref{eq:div} neglects the boundary flux associated with $\bnabla\bcdot(\bstau'\bcdot\bsu')$. 
Here, we justify that approximation. 
The condition $\htau\bcdot\hu=\bs{0}$ is satisfied exactly at the inlets and solid walls, where $\hu=\bs{0}$ is imposed.
At the outlets, we verify that every component of the perturbation velocity and perturbation polymeric stress is smaller than $10^{-5}$ of its corresponding maximum absolute value over the domain, so that the outlet contribution is negligible. This assumption is consistent with the decay of both perturbation fields in sufficiently long outlet channels.

\section{Effects of mesh and polymer--stress diffusivity} \label{appen:mesh}
In \S\,\ref{sec:DNS}, figure~\ref{fig:DQ_eps5e-5_subfig} shows the dependence of the DNS flow solution on three meshes, M1, M2, and M3, when the polymer--stress diffusivity $\epsilon=5\times10^{-5}$. 
Here, we reexamine this dependence in the absence of polymer--stress diffusion, i.e., $\epsilon=0$, as displayed in figure~\ref{fig:DQ_abs_fit_eps0}.  The data reveal qualitatively the same  trend upon varying $\epsilon$. Meanwhile, the critical $Wi$ numbers, $Wi_{cr}=0.3676$ (M1), $0.3638$ (M2) and $0.3616$ (M3), change only weakly, deviating from their counterparts at $\epsilon=5\times10^{-5}$ by less than $1\%$. Despite the small change, the mesh convergence appears more consistent for $\epsilon=0$ than for $\epsilon=5\times 10^{-5}$. This observation suggests that the mesh sensitivity near $Wi_{cr}$ seen in figure~\ref{fig:DQ_eps5e-5_subfig}(a) is likely associated with the finite polymer--stress diffusion.

\begin{figure}[!htbp] 
    \centering
    \includegraphics[width=0.9\linewidth]{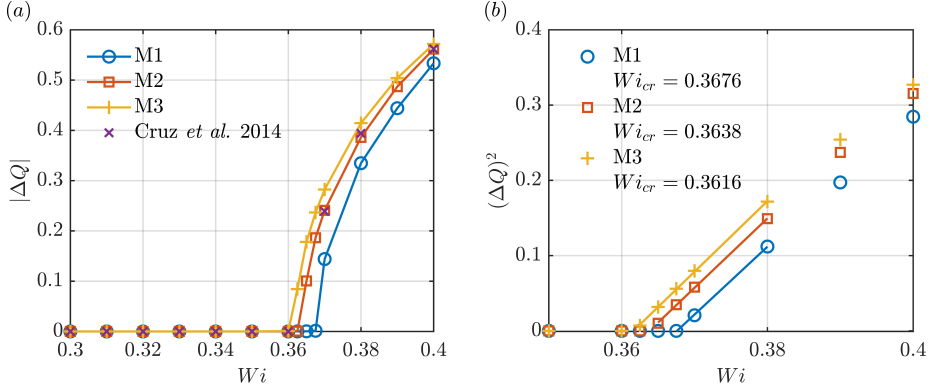}
    \caption{
    Similar to figure~\ref{fig:DQ_eps5e-5_subfig}, while changing $\epsilon=5\times 10^{-5}$ to $\epsilon=0$. Here, the critical Weissenberg numbers slightly change to $Wi_{cr} = 0.3676$, $0.3638$, and $0.3616$ for meshes M1, M2, and M3, respectively.
    }
    \label{fig:DQ_abs_fit_eps0}
\end{figure}

Beyond DNS solutions, we also examine the mesh dependence of the LSA results, as shown in figure~\ref{fig:sigma_eps0_5e-5} (a) for $\epsilon=0$ and (b) $\epsilon = 5\times 10^{-5}$, respectively. In both cases, the temporal growth rate $\sigma_r$ shows satisfactory mesh independence, especially between M2 and M3, in the unstable $Wi$ regime. More pronounced mesh--dependence variations in $\sigma_r$ appear in the stable regime, $Wi<0.35$. However, such variations do not affect identifying the critical value $Wi_{cr}$.

\begin{figure}[!htbp]
    \centering
    \includegraphics[width=0.9\linewidth]{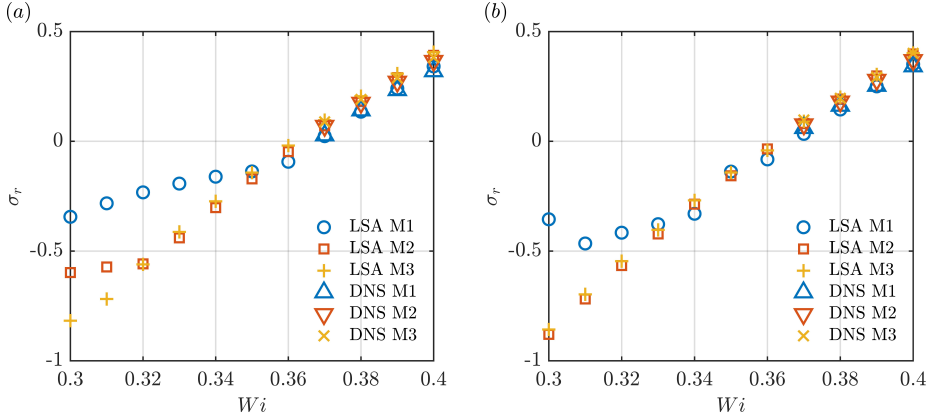}
    \caption{Temporal growth rate $\sigma_r$ versus $Wi$ obtained from LSA and DNS  for three meshes M1, M2, and M3 when $(a)$ $\epsilon = 0$ and $(b)$ $\epsilon = 5 \times 10^{-5}$.     
}
\label{fig:sigma_eps0_5e-5}
\end{figure}

For completeness, the LSA-derived growth rate $\sigma_r$ varying with $Wi\in[0.3, 0.4]$ under four $\epsilon$ values is presented in table~\ref{tab:wi_epsilon_comparison}. Notably, the critical Weissenberg number $Wi_{cr}$ remains within the narrow range $(0.36,0.37)$ for all $\epsilon$ values considered, suggesting that polymer-stress diffusivity does not qualitatively alter the stability characteristics of this flow configuration.

\begin{table}
	\begin{center}
		\def~{\hphantom{0}}
		\begin{tabular}{lcccc}
			$Wi$  & $\epsilon = 0$   & $\epsilon = 10^{-5}$ & $\epsilon = 3 \times 10^{-5}$ & $\epsilon = 5 \times 10^{-5}$ \\ [3pt]
			0.30  & -0.597823 & -0.705707 & -0.884490 & -0.879361 \\
			0.35  & -0.171230 & -0.165617 & -0.160253 & -0.157198 \\
			0.36  & -0.047235 & -0.042049 & -0.037664 & -0.034561 \\
			0.37  & ~0.070937 & ~0.075721 & ~0.079648 & ~0.081772 \\
			0.38  & ~0.183296 & ~0.187649 & ~0.190901 & ~0.192651 \\
			0.39  & ~0.290034 & ~0.294019 & ~0.296542 & ~0.297781 \\
			0.40  & ~0.391572 & ~0.395272 & ~0.397157 & ~0.397909 \\
		\end{tabular}
		\caption{
        Dependence of the LSA-derived growth rate $\sigma_r$ on the dimensionless polymer-stress diffusion $\epsilon$, 
        computed using mesh M2.
        }
		\label{tab:wi_epsilon_comparison}
	\end{center}
\end{table}

\section{Supplementary results}\label{appen:supple}
For completeness, we include several additional figures to support the analysis presented in the main text, with details provided in the corresponding captions.
\begin{figure}[!htbp]
    \centering
    \includegraphics[width=1\linewidth]{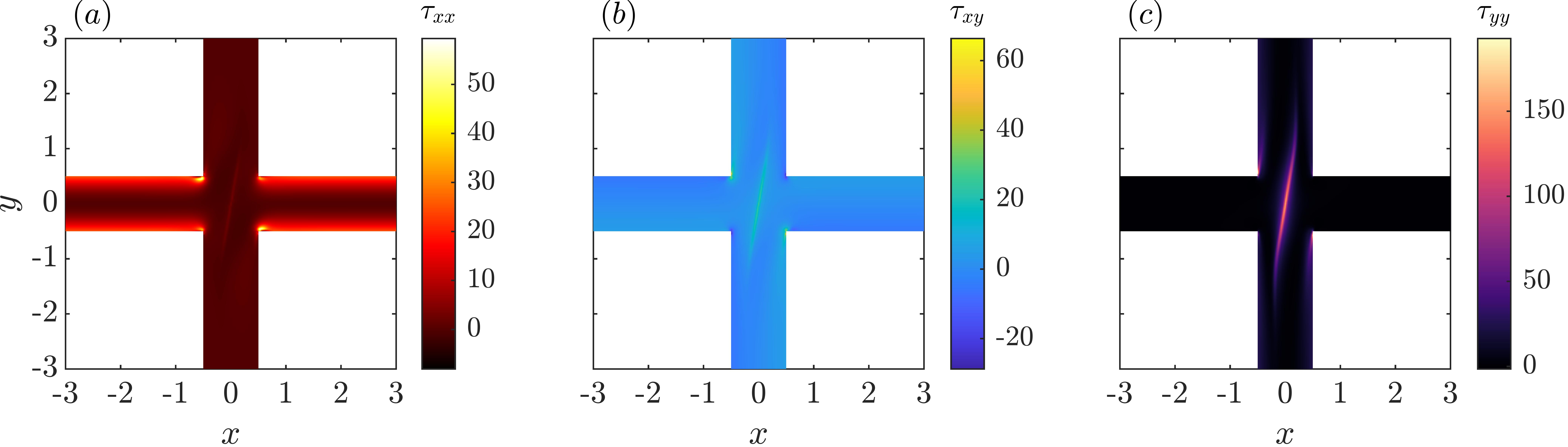}
    \caption{
    Polymeric stress components $\tau_{xx}$ (a), $\tau_{xy}$ (b), and $\tau_{yy}$ (c) of the steady asymmetric DNS solution when $Wi=0.4$ and $\epsilon=5\times10^{-5}$.
    }
    \label{fig:MN40_tau_DNS_Wi0.4}
\end{figure}

\begin{figure}[!htbp]
    \centering
    \includegraphics[width=1\linewidth]{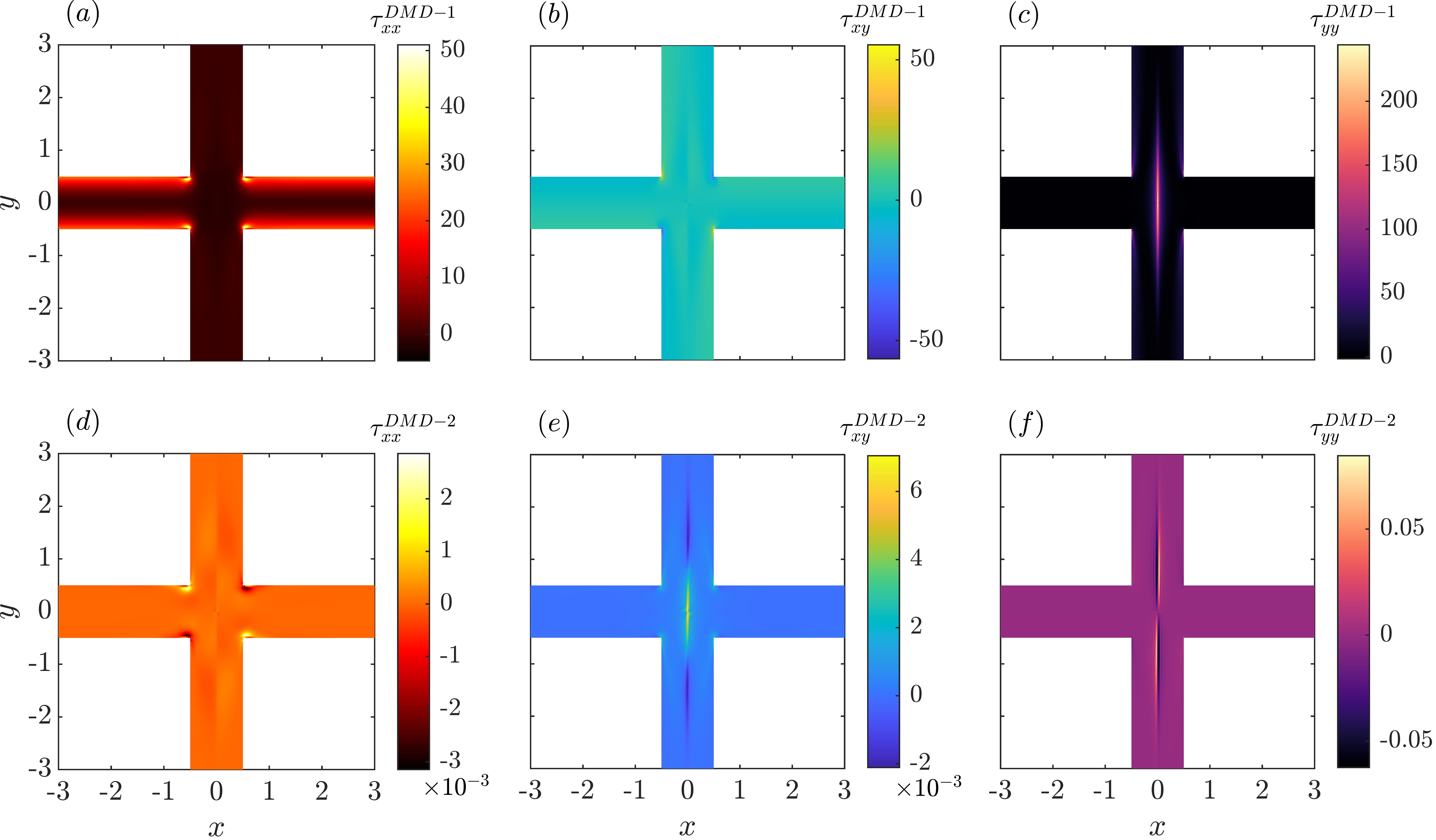}
    \caption{
    Polymeric stress components of two leading DMD modes: (a) $\tau_{xx}^{DMD-1}$, (b) $\tau_{xy}^{DMD-1}$, (c) $\tau_{yy}^{DMD-1}$, (d) $\tau_{xx}^{DMD-2}$, (e) $\tau_{xy}^{DMD-2}$, and (f) $\tau_{yy}^{DMD-2}$.      Here, $Wi=0.4$ and $\epsilon=5e-5$. 
}
    \label{fig:MN40_tau_DMD_mode12_Wi0.4}
\end{figure}

\begin{figure}[!htbp]
    \centering
    \includegraphics[width=1\linewidth]{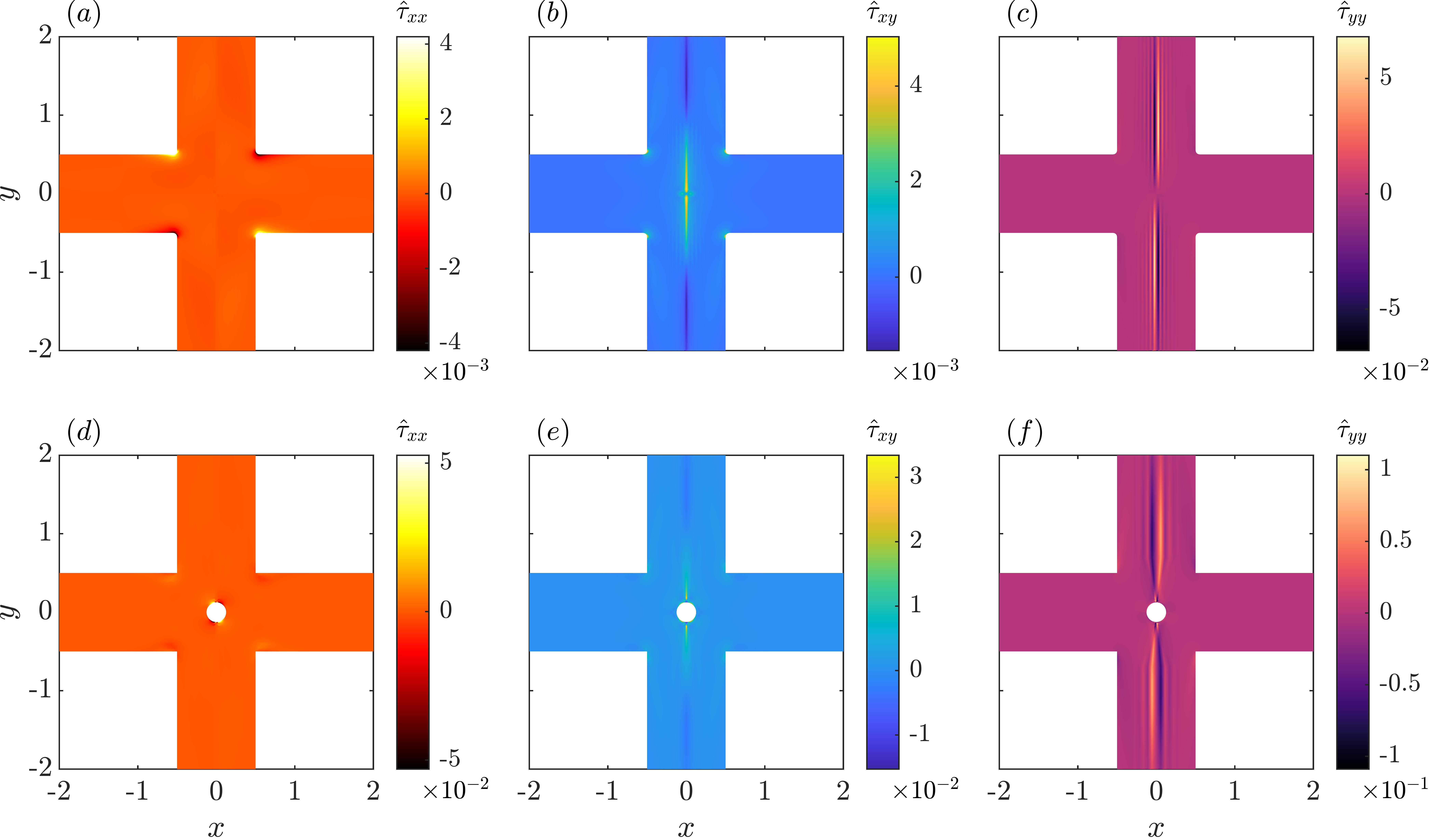}
    \caption{
    (a--c): Polymeric stress components, $\hat{\tau}_{xx}$, $\hat{\tau}_{xy}$, and $\hat{\tau}_{yy}$, of the leading direct eigenmode for the cross-slot channel with rounded corners of arc radius $0.05$ when $Wi=0.4$ and $\epsilon=0$. (d--f): Similar to (a--c), but for the cross-slot channel with a centered cylinder of diameter $0.25$. 
    }
    \label{fig:rc_cyl_eig_tau_2_3}
\end{figure}

\end{appen}\clearpage

%\bibliographystyle{jfm}
%\bibliography{jfm}

\end{document}